\PassOptionsToPackage{pdfpagelabels=false}{hyperref}
\documentclass[useAMS,fleqn,usenatbib]{mnras}
\pdfoutput=1
\usepackage{graphicx}
\usepackage{amsmath,amssymb,amstext,color}
\usepackage[T1]{fontenc}
\usepackage{ae,aecompl}
\usepackage[utf8]{inputenc}
\usepackage{newtxtext,newtxmath}
\usepackage[figure,figure*,table,table*]{hypcap}
\usepackage[dvipsnames]{xcolor}
\usepackage{xparse}
\usepackage[title]{appendix}
\usepackage{chngcntr}
\usepackage{multirow}
\usepackage{caption}
\usepackage{subcaption}
\usepackage{orcidlink}
\usepackage{ulem}

\DeclareRobustCommand{\VAN}[3]{#2}
\let\VANthebibliography\thebibliography
\def\thebibliography{\DeclareRobustCommand{\VAN}[3]{##3}\VANthebibliography}

\input{macros.sty}
\input{hyperlink-year-only-natbib-patch}

\title[Cosmic web effect on galaxies beyond haloes]{The Beyond-Halo Mass Effects of the Cosmic Web Environment on Galaxies}

\author[K. Wang et al.]{
Kuan Wang\orcidlink{0000-0001-7690-2260}$^{1,2}$\thanks{E-mail: kuanwang@umich.edu},
Camille Avestruz\orcidlink{0000-0001-8868-0810}$^{1,2}$\thanks{E-mail: cavestru@umich.edu},
Hong Guo\orcidlink{0000-0003-4936-8247}$^{3}$\thanks{E-mail: guohong@shao.ac.cn},
Wei Wang$^{4}$,
Peng Wang\orcidlink{0000-0003-2504-3835}$^{3,5}$\\
$^{1}$Department of Physics, University of Michigan, Ann Arbor, MI 48109, USA\\
$^{2}$Leinweber Center for Theoretical Physics, University of Michigan, 450 Church St, Ann Arbor, MI 48109, USA \\
$^{3}$Shanghai Astronomical Observatory, Chinese Academy of Sciences, Shanghai 200030, China\\
$^{4}$Purple Mountain Observatory, Chinese Academy of Sciences, Nanjing 210034, China \\
$^{5}$Astronomical Research Center, Shanghai Science \& Technology Museum, Shanghai, 201306, China\\
}

\date{}
\pubyear{2023}

\begin{document}
\label{firstpage}
\pagerange{\pageref{firstpage}--\pageref{lastpage}}
\maketitle

\begin{abstract}
Galaxy properties primarily depend on their host halo mass. Halo mass, in turn, depends on the cosmic web environment.  We explore if the effect of the cosmic web on galaxy properties is entirely transitive via host halo mass, or if the cosmic web has an effect independent of mass.
The secondary galaxy bias, sometimes referred to as ``galaxy assembly bias'', is the beyond-mass component of the galaxy-halo connection.  We investigate the link between the cosmic web environment and the secondary galaxy bias in simulations. We measure the secondary galaxy bias through the following summary statistics: projected two-point correlation function, $\wprp$, and counts-in-cylinders statistics, $\Pncic$.
First, we examine the extent to which the secondary galaxy bias can be accounted for with a measure of the environment as a secondary halo property. We find that the total secondary galaxy bias preferentially places galaxies in more strongly clustered haloes.  In particular, haloes at fixed mass tend to host more galaxies when they are more strongly associated with nodes or filaments. This tendency accounts for a significant portion, but not the entirety, of the total secondary galaxy bias effect.  
Second, we quantify how the secondary galaxy bias behaves differently depending on the host halo proximity to nodes and filaments.  We find that the total secondary galaxy bias is relatively stronger in haloes more associated with nodes or filaments.
We emphasise the importance of removing halo mass effects when considering the cosmic web environment as a factor in the galaxy-halo connection.
\end{abstract}



\section{Introduction}
\label{sec:intro}

Numerical simulations and galaxy surveys have shown that the large-scale structure of the Universe can be described by an intricate network of voids, sheets, filaments, and nodes, which is known as the \emph{cosmic web} \citep{joeveer1978,delapparent86,bond1996}. The cosmic web originates from primordial density fluctuations and evolves under gravitational interactions, creating a variety of cosmic environments \citep[see][and references therein]{bond_strauss_cen2010,cautun2014,wang2024_fil_bound}. In general, matter tends to flow out of voids and onto surrounding sheets, and accrete through filaments into nodes. On smaller, nonlinear scales, virialised dark matter haloes populate the cosmic web, and galaxies form and evolve in the potential wells of these haloes \citep[see, e.g.,][]{mo_vdb_white10}. While voids dominate the volume of the Universe, filaments and nodes contain most of the mass, as well as haloes and galaxies \citep{pimbblet2004,aragon-calvo2010}.

Tidal forces from the environment affect dark matter haloes. Depending on the type of environment, haloes experience different tidal effects and display different assembly characteristics \citep[e.g.,][]{gottlober2001,jing2007,hahn2009,paranjape18}. Studies have revealed that the formation time, spin, concentration, and shape of haloes are related to their position in the cosmic web, as measured by their distances to neighbouring structures and / or local densities \citep[e.g.,][]{sheth_tormen2004,wechsler06,hahn2007a,wang_mo_jing_yang_wang2011,chira2021}. For example, halo shapes tend to align with neighbouring sheets and / or filaments, leading to alignments between haloes, while halo spins have a mass-dependent tendency to be parallel or perpendicular to their parent structure \citep[e.g.,][]{kasun_evrard2005,hahn2007b,zhang2009,trowland2013,forero-romero2014}. More recent work \citep[e.g.,][]{borzyszkowski2017,tojeiro2017,yang2017,musso2018,ramakrishnan2019} has also shown that the cosmic web environment has an influence on \emph{halo assembly bias}, the dependence of halo clustering on halo properties other than mass \citep{gao_etal05,gao_white07,li_etal08}.

Haloes are the main drivers of galaxy formation and evolution \citep{whiterees78,blumenthal_etal84}. By modelling the statistical relationship between galaxy properties and the properties of their host haloes, we can interpret cosmological observations \citep[e.g.,][]{zehavi_etal11,guo2015,vakili_2019,lange2018,wechsler_tinker18}. One of the simplest forms of this galaxy-halo connection assumes that the mass of a halo completely determines the characteristics of the galaxies it contains \citep[e.g.,][]{zheng07}. However, this mass-only assumption is insufficient for precision cosmology \citep[e.g.,][]{wu08,zentner_etal14,McCarthy_2018}. Subsequently, more recent galaxy-halo models include an additional dependence of galaxy properties on secondary halo properties at a fixed halo mass \citep[e.g.,][]{hearin_etal16,Lehmann2017}.

We can differentiate between the internal and environmental halo properties. The connection between galaxy properties and internal halo properties, such as halo concentration, is known as \emph{galaxy assembly bias} \citep[e.g.,][]{croton_etal07}. Galaxy assembly bias has an effect on galaxy clustering, which can be detected in observations \citep[e.g.,][]{cooper2010,wang_etal13,zentner_etal19}. On the other hand, environmental halo properties, such as matter density on intermediate scales, are naturally linked to halo clustering. Any dependence of galaxy properties on these environmental halo properties will be reflected in galaxy clustering as well \citep[e.g.,][]{Artale_2018,zehavi2018,Xu2021}. Since internal and environmental halo properties are usually correlated, these two types of dependencies are also connected. Following the ideas of \citet{mao_etal18}, we suggest the use of the term \emph{secondary galaxy bias} (SGB) to refer to all dependencies of galaxy properties on internal or environmental halo properties at a fixed halo mass.

Since the cosmic web has a major impact on dark matter haloes, it is reasonable to expect that the cosmic web also plays a role in shaping galaxy properties. In fact, studies have demonstrated that star formation, colour, morphology, and stellar mass are all strongly and non-trivially dependent on the type and density of the environment \citep[e.g.,][]{dressler80,kodama2001,blanton05,gonzalez2009,sobral2011,eardley2015,kraljic2018,alam2019,aragon-calvo2019}. Additionally, there is evidence of statistical alignments between galaxies and their large-scale environment \citep[e.g.,][]{sales_lambas2004,azzaro2007,faltenbacher2009,hahn2010,zhang2013}. Furthermore, both numerical and observational studies have confirmed the correlation between galaxy spins and the environment \citep{navarro2004,paz2008,tempel2013a,tempel2013b}, which is caused by tidal torques \citep{efstathiou1979,white1984}.

We note that the majority of research on the relationship between galaxies and their environment does {\it not} differentiate between the environmental effect and the halo mass effect. This leads us to ask: {\it Does the galaxy-halo connection have a component driven by the cosmic web independent of halo mass?} We consider two aspects: (i) how much of the total SGB can be attributed to the cosmic web environment as a secondary halo property; and (ii) whether the SGB effect behaves differently in different cosmic web environments.

In this paper, we utilise the \textsc{IllustrisTNG} hydrodynamical simulation \citep[e.g.,][]{pillepich2018b_TNG,nelson2019a_TNG} to explore the connection between the SGB and the cosmic web. We quantify the environment of galaxies by measuring their proximity to nodes or filaments in the cosmic web, identified using the \texttt{DisPerSE} cosmic web finder \citep{sousbie2011I,sousbie2011II}. We measure the strength of the SGB effect using the shuffling procedure developed in \citet{croton_etal07} \citep[see also][for recent applications of this technique]{McCarthy_2018,Xu2021,yuan_etal22}. Our findings shed light on the relationship between the secondary galaxy bias and the cosmic web environment, thereby helping to elucidate the physics of galaxy formation and evolution in the context of the large-scale structure.

This paper is organised as follows.
In \autoref{sec:sim_methods}, we introduce the data set and methods that we use in our analyses.
In \autoref{sec:hod_measure}, we examine the dependence of the directly measured halo occupation distribution on different cosmic environments.
In \autoref{sec:sgb_denv}, we treat the cosmic web environment as a secondary halo property, and study its contribution to the total secondary galaxy bias.
In \autoref{sec:sgb_split}, we study how the secondary galaxy bias differs in different cosmic web environments.
We discuss our findings in \autoref{sec:discussion} and draw conclusions in \autoref{sec:conclusion}.
\autoref{sec:appendixA}, \autoref{sec:appendixB}, \autoref{sec:appendixC} and \autoref{sec:appendixD} describe additional tests.

\section{Data and Methods}
\label{sec:sim_methods}

\subsection{\textsc{IllustrisTNG} simulation}
\label{sec:sim}

This work utilises the TNG300-1 run of the IllustrisTNG simulation suite \citep{marinacci2018_TNG,naiman2018_TNG,nelson2018a_TNG,pillepich2018b_TNG,springel2018_TNG,nelson2019a_TNG}, which is a set of large-scale, cosmological, gravo-magnetohydrodynamical simulations conducted with the \texttt{AREPO} code \citep{springel_2010}. The simulations are based on the Planck 2015 cosmology \citep{planck2016}, with $\Omega_{\Lambda,0} = 0.6911$, $\Omega_{m,0} = 0.3089$, $\Omega_{b,0} = 0.0486$, $\sigma_8 = 0.8159$, $n_s = 0.9667$ and $h = 0.6774$. The TNG300-1 run is the high-resolution full-physics run with the largest volume, having a box size of $L_{\rm box} = 205\Mpch$ and dark matter and baryon mass resolution of $4\times 10^7\Msunh$ and $7.6\times 10^6\Msunh$, respectively.

In the TNG simulation, the haloes are identified using the standard friends-of-friends (FoF) algorithm \citep[e.g.,][]{davis85_fof}.
In our analyses, we adopt the virial masses of the haloes, which are taken from the group catalogue.
However, we test that our results are not qualitatively dependent on the mass definition, with alternative halo mass definitions in \autoref{sec:appendixC}.
Subhaloes, which contain individual galaxies, are identified with the \texttt{Subfind} algorithm \citep{subfind}. The stellar masses and positions of the galaxies are obtained from \texttt{Subfind}, and we focus on galaxies with stellar masses higher than $10^8\Msun$ in this paper, unless otherwise specified.

\subsection{Cosmic web classification}
\label{sec:classify}

We use the \texttt{DisPerSE} cosmic web finder \citep{sousbie2011I,sousbie2011II} to identify structures in the simulation volume. \texttt{DisPerSE} provides automatic identification of topological structures such as nodes, filaments, walls and voids, namely the cosmic web, based on the Discrete Morse theory \citep[e.g.,][]{forman2002user}. \texttt{DisPerSE} uses discrete distributions of particles in simulations or sparse observational catalogues to estimate a density field. Nodes are critical points in the density field, with filaments being the unique integral lines connecting them. Saddle points are minima along filaments. In this study, under the consideration of the mass resolution of the simulation, sample size and be able to compare with the observational data, we choose galaxies with stellar masses above $10^{8.5}\Msunh$ as the input tracers of \texttt{DisPerSE} for the cosmic web searching. Our application of the \texttt{DisPerSE} algorithm is the same as that of \cite{galrraga2020}, setting the signal to noise of $3\sigma$ as the criterion to identify filaments, and we verify that the results are quite similar to their catalogues.  We obtained a total of 11,446 filaments, and the distance of each galaxy to its nearest filament or node is recorded.

\begin{figure*}
    \centering
    \includegraphics[width=\textwidth]{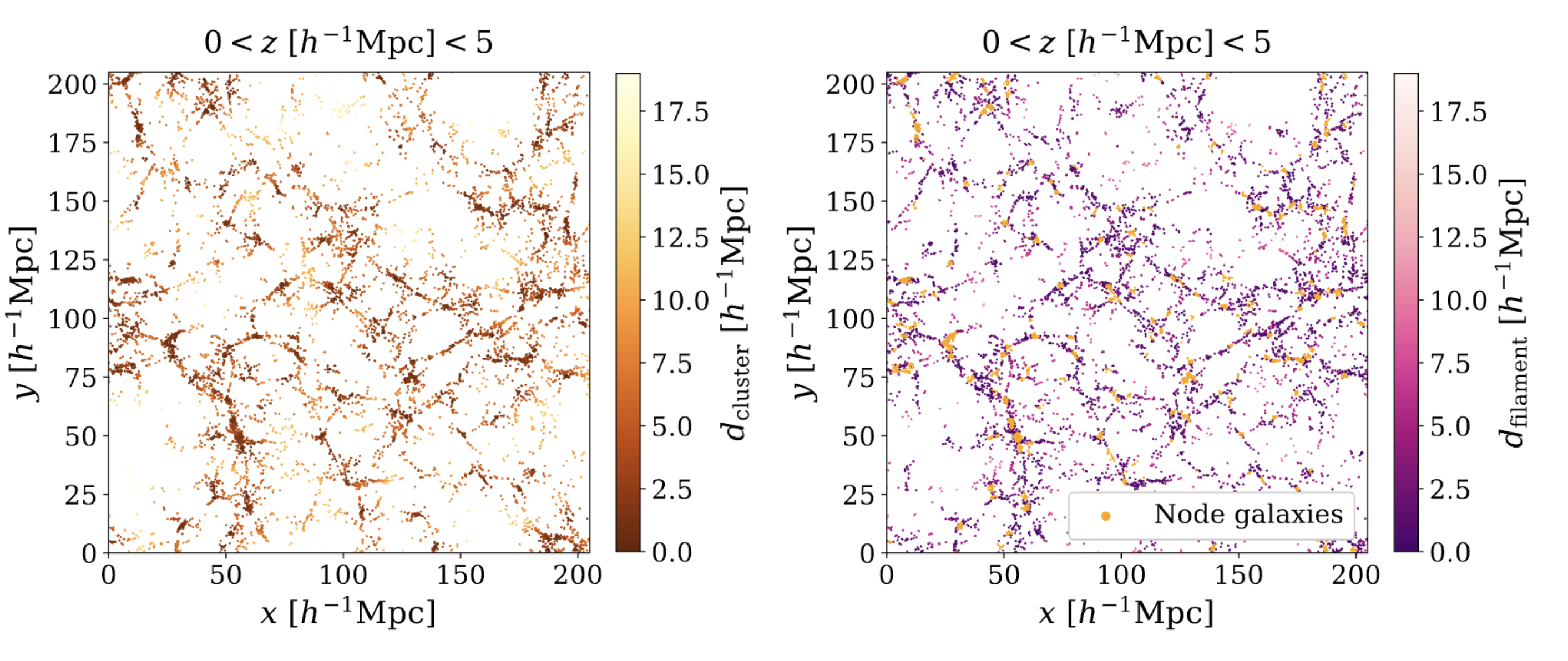}
    \caption{In this figure we show a thin slice of the simulation projected in the $z$ direction. 
    We plot the spatial distribution of galaxies with stellar masses above $10^8\Msun$, where each scatter point represents a galaxy. Left panel shows galaxies colour coded by distance to the nearest node, $\dnode$. Right panel shows galaxies colour coded by distance to the nearest filament, $\dfil$.
    Orange dots in the right panel show ``node galaxies'' (with host haloes within $2\Mpch$ of the nearest node).
    We exclude node galaxies from our analysis with $\dfil$ to isolate the impact of filaments, because neighbouring nodes dominate their environment. }
    \label{fig:gal_scatter_dis}
\end{figure*}

\begin{figure*}
    \centering
    \includegraphics[width=\textwidth]{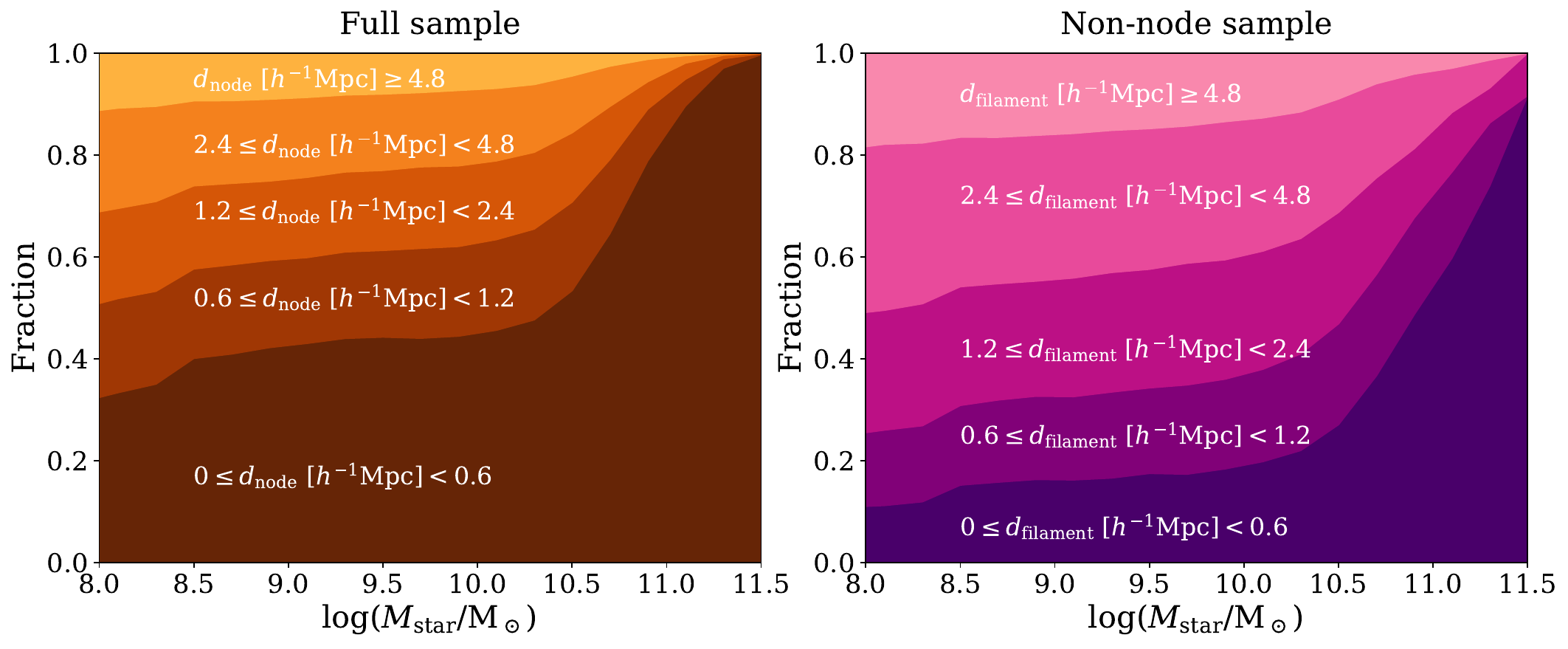}
    \caption{In this figure we show the respective fractions of galaxies in different distance bins. Left panel shows distances to nodes, $\dnode$, as a function of the galaxy stellar mass $\Mstar$.  Right panel shows distances to filaments, $\dfil$, as functions of stellar mass. We label the $\dnode$ and $\dfil$ bins within the figure.
    The right panel excludes all node galaxies to isolate the impact of the filament environment.}
    \label{fig:gal_fraction_dis}
\end{figure*}

We quantify the cosmic web environment of galaxies in terms of their proximity to nearby dense structures, namely, the distance to the nearest node, $\dnode$, and the nearest filament, $\dfil$. We investigate the effects of nodes and filaments separately. Galaxies that are close to nodes are mainly affected by the node environment and are not as sensitive to the filaments around them, and in our analyses of $\dfil$, we exclude galaxies with $\dnode<2\Mpch$. We refer to this sample as the \emph{non-node sample}, in contrast to the \emph{full sample}, which includes all galaxies with stellar masses greater than $10^8\Msun$. In \autoref{fig:gal_scatter_dis}, we illustrate these distances with a scatter plot of galaxies in a thin slice of the simulation box. We observe that most galaxies are distributed around nodes and along filaments, forming a web-like structure, as expected. In \autoref{fig:gal_fraction_dis}, we show the fractions of galaxies with different distances to nodes and filaments, as functions of galaxy stellar mass. It is clear from the figure that more massive, brighter galaxies tend to inhabit node and filament environments.
We note that apart from a small number of haloes that reside in nodes and filaments, the vast majority of haloes have radii that are small compared to their distances to the closest node or filament.
The distance proxies do not capture all the information from the cosmic web environment, and we will discuss other possibilities in \autoref{sec:conclusion}.

\subsection{Statistics}
\label{sec:stats}

The connection between galaxies and haloes, together with the halo distribution, determines the spatial distribution of galaxies, and we measure galaxy clustering through summary statistics.
We select two distinct statistics, the projected two-point correlation function $\wprp$ and the counts-in-cylinders statistic $\Pncic$, both of which are based on finding pairs of galaxies. We use the real-space positions of galaxies from the simulation, disregarding peculiar velocities. To make it easier to compare our results with those from observations, which will be explored in our future work, we still use projected statistics, which are less affected by redshift space uncertainties. We take the $z$-axis as the line-of-sight direction for our measurements. We use the \texttt{halotools} package \citep{halotools} to make our measurements.

Two-point correlation functions encode the majority of information in near-Gaussian fields and are used as standard statistics in the literature.
We measure the projected two-point correlation function,
\begin{equation}
\label{eq:wp}
\wprp = 2\int_{0}^{\pimax}d\pi \ \xi(\rp,\pi),
\end{equation}
where $\xi(\rp,\pi)$ is the excess probability of finding galaxy pairs with projected and line-of-sight separations $\rp$ and $\pi$, respectively.
We choose $\pimax=40\Mpch$, and compute $\wprp$ in 10 logarithmically spaced radial bins between $\rp=0.1\Mpch$ and $\rp=31.6\Mpch$.

\begin{figure}
    \centering
    \includegraphics[scale=0.38, trim={0.5cm 6.3cm 0.5cm 0cm}, clip]{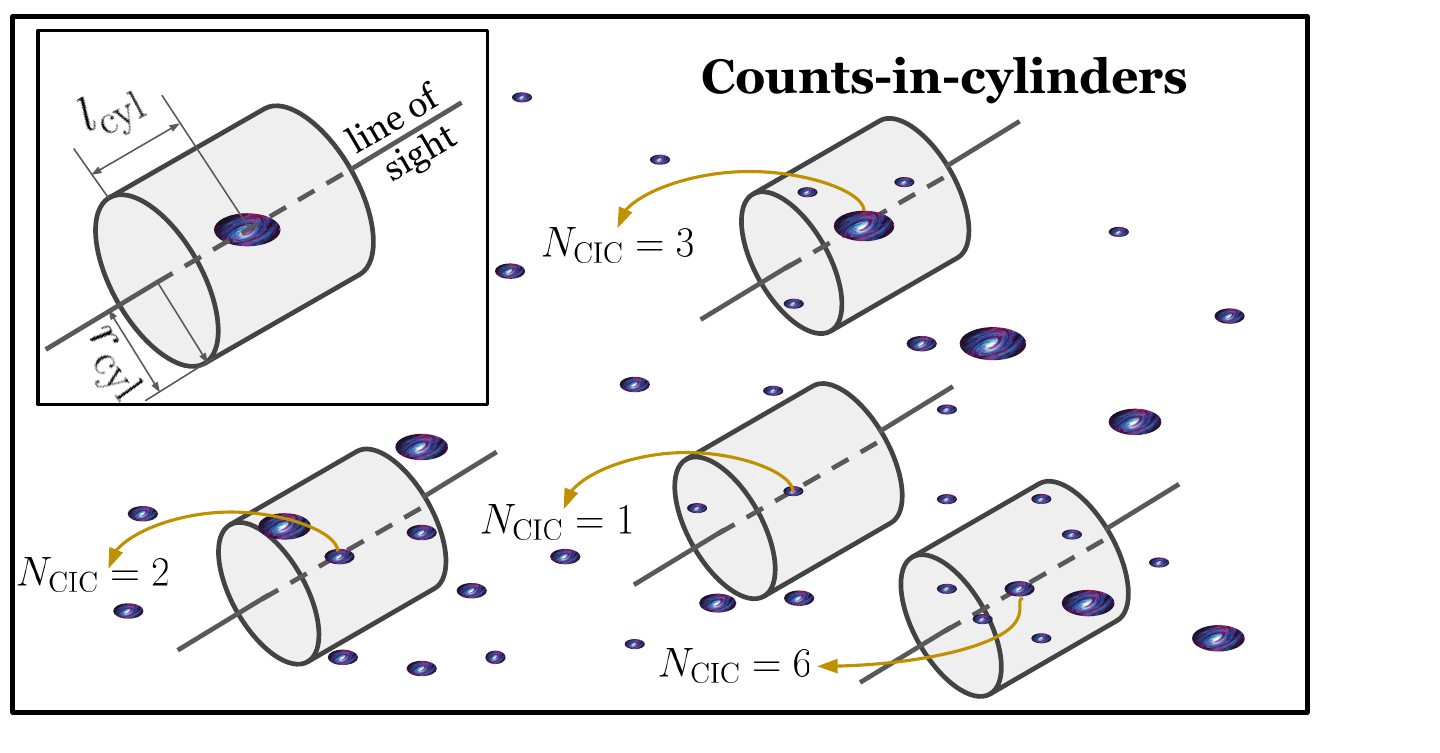}
    \caption{In this figure, we illustrate the definition of the counts-in-cylinders statistics.
    The cylinder is placed around each galaxy in the sample along the line of sight, and has a radius of $\rcyl$ and half-length of $\lcyl$.
    The number of companions that fall in the cylinder is then counted for each galaxy.   Counts-in-cylinders provide sensitivity to higher order statistics and to less dense regions of the galaxy distribution, both of which are complementary to information in the two-point correlation function.}
    \label{fig:CIC_cartoon}
\end{figure}

In \citet{wang2019_CIC,wang2022_SDSS_CIC}, we have demonstrated that the counts-in-cylinders statistics are an informative complement to the two-point statistics, because they encode higher-order information of the galaxy field. We measure the counts-in-cylinders statistic for a sample of galaxies by constructing cylinders of radius $r_{\rm cyl}$ and half-length $l_{\rm cyl}$ along the line of sight, centring them on each galaxy in the sample. We then count the number of companion galaxies in the sample that fall into each cylinder and use the distribution of this companion count as our summary statistic of the galaxy spatial distribution (illustrated in \autoref{fig:CIC_cartoon}). In real space, we use the same $r_{\rm cyl}$ and $l_{\rm cyl}$ of $5\Mpch$ to probe sufficiently large scales in the spatial distribution of galaxies, and denote the count statistics with this cylinder size as $P_5(\Ncic)$. We have tested that different cylinder sizes do not affect our qualitative results (see \autoref{sec:appendixB}). 

\subsection{Catalogue shuffling}
\label{sec:shuffle}

We measure the strength of the SGB with the shuffling technique, first proposed by \citet{croton_etal07}. Here, we present a detailed description of the shuffling method, which is also illustrated in \autoref{fig:shuffling_cartoon}.

\subsubsection{Mass shuffling}
\label{sec:mass_shuffle}
\begin{figure}
    \centering
    \includegraphics[scale=0.38, trim={0.6cm 0.5cm 0 0cm}, clip]{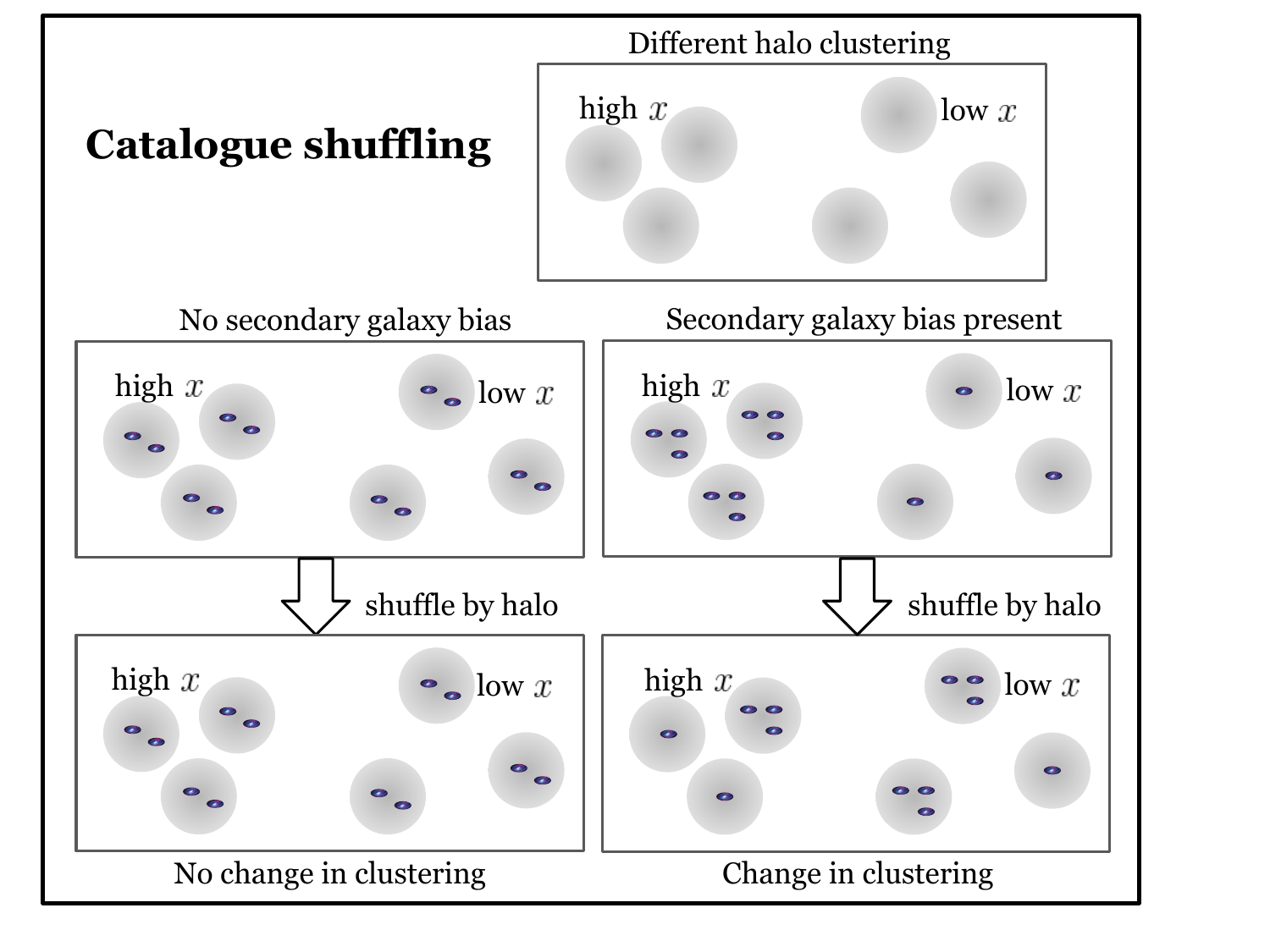}
    \caption{In this figure, we explain the catalogue shuffling technique with which we measure the strength of secondary galaxy bias (SGB), where $x$ is some secondary halo property.
    We schematically demonstrate the effect of shuffling galaxies between haloes with the same masses, and compare between cases with and without SGB.
    A detailed account of the method can be found in \autoref{sec:shuffle}.}
    \label{fig:shuffling_cartoon}
\end{figure}

The SGB is the dependence of galaxy occupation on some secondary halo property (denoted as $x$ in \autoref{fig:shuffling_cartoon}), which is either internal or environmental. This effect can be detected in galaxy clustering through its combination with the underlying halo clustering. To isolate the SGB, galaxies are randomly shuffled among haloes of the same mass, while preserving the phase-space distribution of satellite galaxies with respect to the central galaxy. This erases any dependence of galaxy occupation on halo properties other than mass. Comparing the shuffled and original clustering provides a quantification of the SGB. If there is no SGB present, the shuffling has no impact on the measured galaxy clustering. However, if there is SGB in the sample, the shuffling alters the galaxy clustering. In practise, galaxies are shuffled among haloes in narrow mass bins of 0.1 dex, over which the scatter introduced by the mass dependence is typically small.

\subsubsection{Double shuffling}
\label{sec:double_shuffle}

It is possible to further investigate the origins of the SGB with the double shuffling technique. This technique fixes the halo mass and a secondary halo property when reassigning galaxies among haloes. This eliminates any dependence of galaxy occupation on halo properties other than mass and the fixed secondary halo property. By comparing the mass shuffles, the double shuffles, and the original galaxy distribution, it is possible to determine the portion of the SGB that can be attributed to the secondary halo property, such as $\dnode$ or $\dfil$, and the portion that cannot, thus indicating its relative importance in determining galaxy occupation.
We bin the secondary property, in our case the distance from the closest node or filament, in 10 percentile bins.
The number of objects in each bin varies, but is typically above 1000, allowing for sufficient statistics for the shuffling.
    
\subsection{Secondary galaxy bias strength}
\label{sec:strength}

\subsubsection{Ratio measurement}
\label{sec:ratio}

We measure the strength of the secondary galaxy bias (SGB) by comparing the statistics of the original and shuffled galaxy distributions, $\mathbf{d}_{\rm 1} / \mathbf{d}_{\rm 2}$, where $\mathbf{d}$ is the data vector. The deviation of these ratios from 1 can be used to detect the SGB, as the difference reflects the effect of the SGB on the statistics. For instance, ratios greater than 1 indicate that the SGB present in the original sample increases the value of the statistics, and vice versa. When $\mathbf{d}_{\rm 1}$ and $\mathbf{d}_{\rm 2}$ are measured from the original sample and mass shuffle, respectively, $\mathbf{d}_{\rm 1} / \mathbf{d}_{\rm 2}$ reveals the full extent of SGB from all sources; when $\mathbf{d}_{\rm 1}$ and $\mathbf{d}_{\rm 2}$ are measured from the double shuffle and the mass shuffle, respectively, $\mathbf{d}_{\rm 1} / \mathbf{d}_{\rm 2}$ indicates the SGB that can be attributed to the second halo property.

\subsubsection{Uncertainty estimation}
\label{sec:errorbar}

We provide an estimate of the statistical significance of our SGB signal by computing jackknife uncertainties of the ratios, as was done in \citet{hadzhiyska_etal21}. We divide the original simulation box and the shuffled simulation box into $5\times5$ cuboid cells, each of size $41\ \Mpch \times 41\ \Mpch \times 205\ \Mpch$. The long axis of each cuboid is the same as the length of the simulation box and is assumed to lie along the line of sight. We calculate the data vector for the jackknife subsamples, excluding one cuboid at a time. The jackknifed ratios are calculated between pairs of jackknife subsamples that exclude the same cuboid, so that the ratios are only dependent on the changes in the galaxy occupation of haloes, not differences in host haloes themselves. We use the jackknife errors from these ratios to represent the uncertainty.

\section{Measured Halo Occupation Distribution}
\label{sec:hod_measure}

\begin{figure*}
    \centering
    \includegraphics[width=0.9\textwidth, trim=0.5cm 1.5cm 0.5cm 2.5cm]{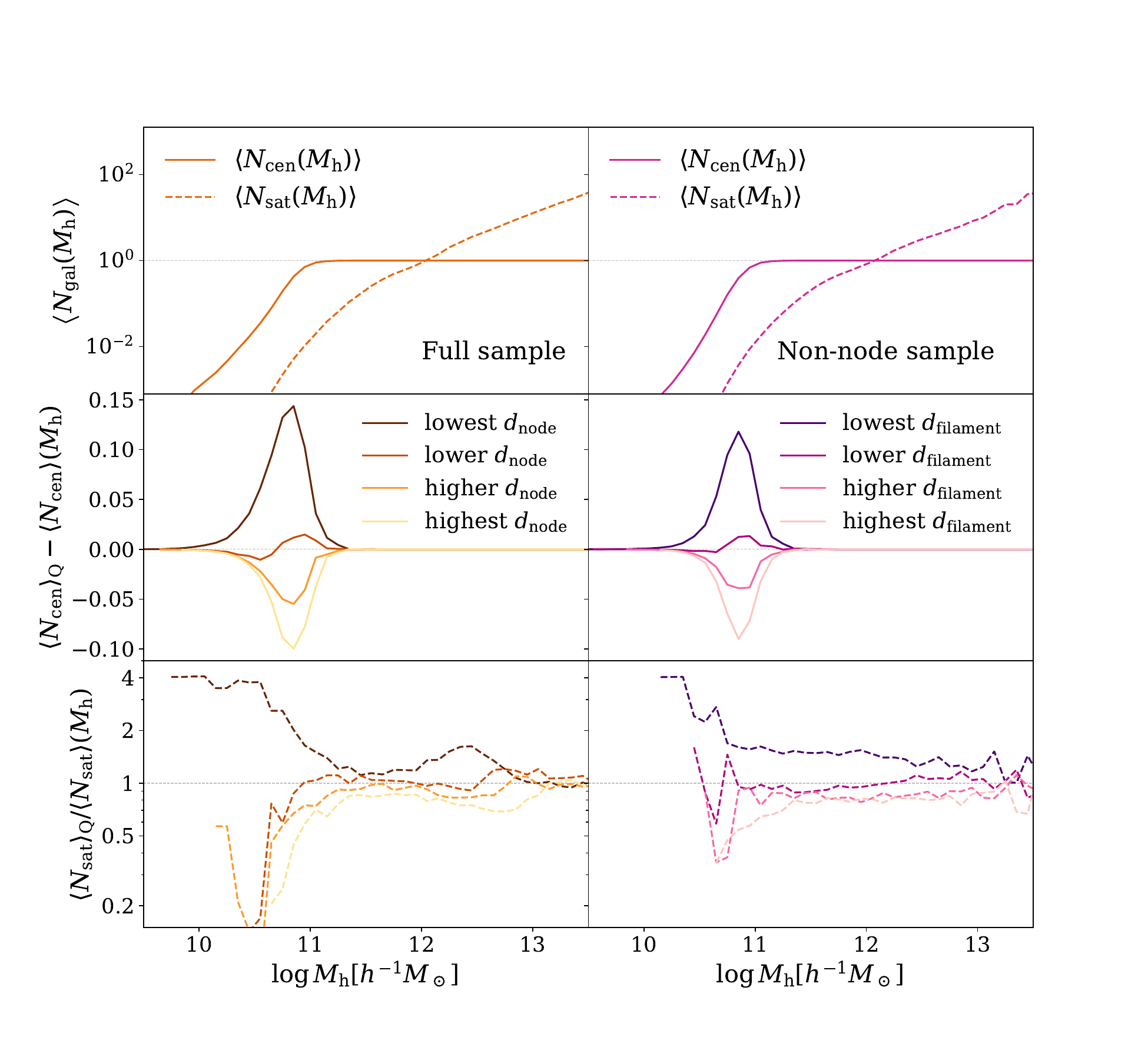}
    \caption{In this figure, we present the HOD measured from the full sample (left column) and the non-node sample (right column).
    In each column, the top panel shows the mean number of central and satellite galaxies in the entire sample as functions of halo mass, in solid and dashed curves respectively, as labelled in the panels.
    The middle panel shows the dependence of the central galaxy occupation on $\dnode$ or $\dfil$ at fixed halo masses.
    We plot the dependence in terms of the difference of the central galaxy occupation between each $\dnode$ or $\dfil$ quartile, $\left<N_{\rm cen}(M_{\rm h})\right>_Q$, and the entire sample, $\left<N_{\rm cen}(M_{\rm h})\right>$.
    Line colours correspond to different quartiles, labelled in the middle panels.
    Similarly, the bottom panel shows the dependence for the satellite occupation, in terms of $\left<N_{\rm sat}\right>_Q/\left<N_{\rm sat}\right>(M_{\rm h})$, and in logarithmic scale. Both central and satellite galaxies preferentially populate haloes that are more strongly associated with either nodes or filaments.}
    \label{fig:hod}
\end{figure*}

The Halo Occupation Distribution (HOD) \citep[e.g.,][]{berlind02,kravtsov04a,zheng07} is a commonly used technique for modelling the relationship between galaxies and haloes. In its simplest form, the HOD assumes that the number of galaxies in a halo is determined solely by the halo mass. Secondary galaxy bias, however, implies a violation of this assumption. The HOD depends on the selection of the galaxy sample, as galaxies of different properties inhabit haloes in different ways. In this section, we will look at the HODs measured from the galaxy samples in \textsc{IllustrisTNG} and investigate whether and how they vary depending on the environment.

\subsection{HODs of the entire samples}
\label{sec:hod_all}

The HODs of central and satellite galaxies are usually modelled separately since dark matter haloes acquire them in distinct ways. In the top row of \autoref{fig:hod}, we display the central and satellite occupation as a function of halo mass, for both the full sample and the non-node sample. These are calculated from the ratio of galaxy number to halo number within each narrow halo mass bin, in a nonparametric form. The measured HODs are in agreement with expectations. Each halo can have either 0 or 1 central galaxy, but any number of satellite galaxies, and more massive haloes host more galaxies. The occupation for the non-node sample is limited to lower halo masses, since the most massive haloes and their galaxies are excluded.

\subsection{Environmental dependence of the HOD}
\label{sec:hod_env}

If the properties of galaxies depend not only on the mass of the halo, but also on a secondary halo property, $x$ (in this case, $\dnode$ or $\dfil$), haloes of the same mass with different values of $x$ will have different numbers of galaxies in a given sample. We divide the haloes in each narrow mass bin into four quartiles of $\dnode$ ($\dfil$), and by comparing the HODs of the quartiles, the SGB associated with $\dnode$ ($\dfil$) can be determined.

We compare the occupation of the central galaxy for each quartile with respect to the entire sample in the middle row of \autoref{fig:hod}. We find that haloes closer to dense structures are more likely to host a central galaxy, with the preference being stronger for the full sample with nodes. The mean central number transitions from 0 at low halo masses to 1 at high halo masses, and the difference between quartiles is most prominent at masses slightly below $10^{11}\Msunh$. Furthermore, the dependence on node (filament) proximity is stronger at lower $\dnode$ ($\dfil$), indicating that nodes and filaments mostly affect their immediate surroundings, and environments become less distinct from each other when they are far from these dense structures.

In the bottom row of \autoref{fig:hod}, we compare the mean satellite occupation of each quartile to that of the entire sample, $\left<N_{\rm sat}\right>_Q/\left<N_{\rm sat}\right>$, as a function of the halo mass, $M_{\rm h}$. The satellite statistics are noisy due to the low number densities of massive objects at the high mass end and the incomplete halo sample at the low mass end, caused by the finite-mass resolution of the simulation. Nevertheless, we still observe a trend similar to the central occupation, where haloes closer to nodes or filaments tend to host more satellite haloes, with the dependence weakening as the distance increases. This measurement suggests that the effects of nodes and filaments on satellite occupation are comparable.

In conclusion, we have demonstrated that there is a greater presence of node-related SGB than filament-related SGB in the central galaxy component of the TNG galaxy sample. Nevertheless, this method is restricted to individual secondary halo properties and cannot measure the total SGB from all sources. To address this, we will use the shuffling technique to determine the relative contribution of the environment-related SGB to the total SGB in the following sections.

\section{Environment as Secondary Halo Property}
\label{sec:sgb_denv}

We seek to answer two questions in this section: (i) what is the total amount of secondary galaxy bias (SGB) present in the IllustrisTNG galaxy sample, and (ii) how much of the total SGB can be attributed to the environmental properties, as quantified by the distances to nodes and filaments, $\dnode$ and $\dfil$. To answer the first question, we compare $\wprp$ and $\Pncic$ measurements between the original galaxy sample and the mass shuffled sample. To answer the second question, we compare measurements from the sample shuffled by both halo mass and the environment property, $\dnode$ or $\dfil$, and the sample shuffled by halo mass alone.

\subsection{Original measurements}
\label{sec:stat_orig}

\begin{figure*}
    \centering
    \includegraphics[width=0.9\textwidth, trim={10 0 0 0}, clip]{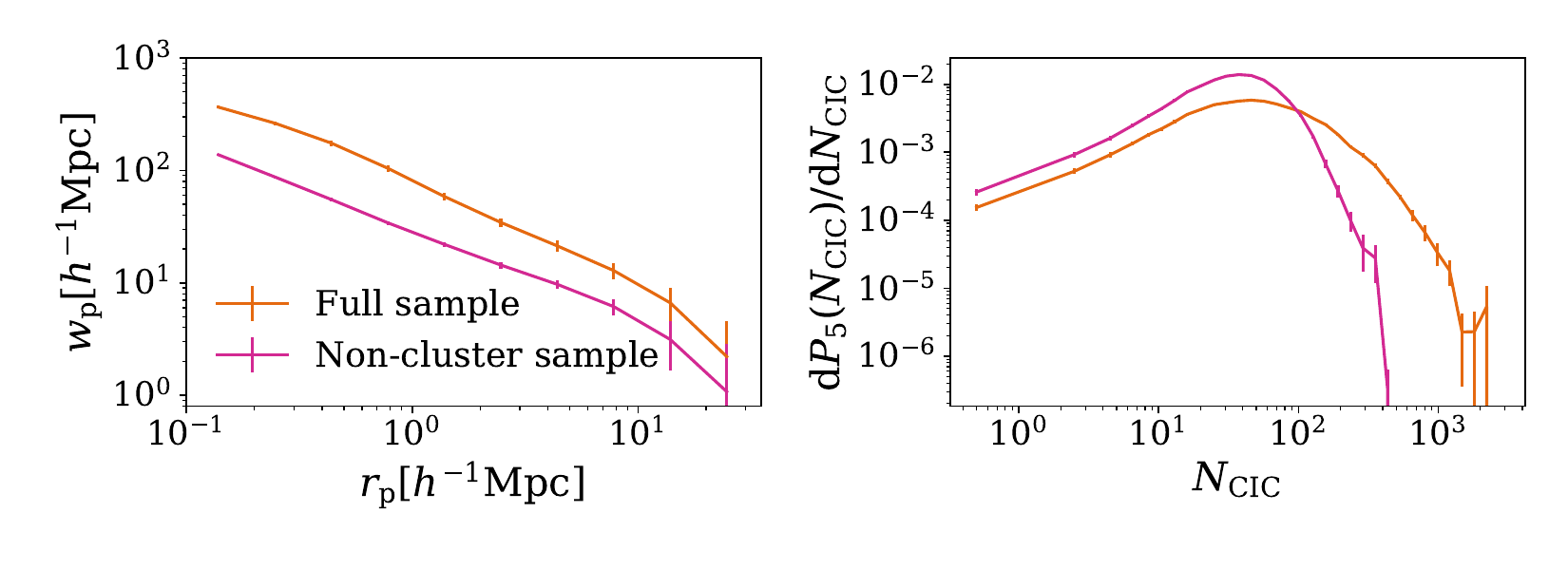}
    \caption{In this figure we present the measurements of the projected two-point correlation function $\wprp$ (left panel) and counts-in-cylinders statistics $\Pncic$ with a cylinder size of 5$\Mpch$ (right panel).
    Orange lines indicate measurements from the full galaxy sample for the $\dnode$ analysis and magenta lines indicate measurements from the non-node sample for the $\dfil$ analysis, as labelled in the left panel.
    The jackknife errors on the measurements are shown as error bars. 
    For some data points, the sizes of the error bars are barely visible.
    We plot the projected two-point correlation function $\wp$ as a function of the projected separation $\rp$.
    The counts-in-cylinders statistics are represented as the probability distribution of the number of companions $\Ncic$, normalised by the bin widths.
    The exclusion of node galaxies suppresses the two-point correlation function and reduces the fraction of groups with higher companions.}
    \label{fig:stat_orig}
\end{figure*}

We first measure the statistics of the original galaxy sample from TNG300-1.
In \autoref{fig:stat_orig}, we show the measurements for both the full and non-node samples, with $\wprp$ in the left panel and $\Pncic$ in the right panel.
The error bars represent jackknife errors, estimated using the same subsampling scheme as described in \autoref{sec:errorbar}.
The two-point clustering is significantly reduced when node galaxies are excluded, and the fraction of groups with higher companion counts also decreases, resulting in a higher probability of having fewer companion galaxies in a cylinder, which is in line with our expectation, as node galaxies are a major contributor to the abundance of pairs and neighbours.

\subsection{Secondary galaxy bias signal}
\label{sec:sgb_signal_all}

We now repeat the measurements for different galaxy catalogues and compare the results between the original and shuffled samples. We make four sets of comparisons: 
\begin{enumerate} 
\item The original full sample versus the full sample shuffled by mass; 
\item The full sample shuffled by mass and $\dnode$ versus the full sample shuffled by mass; 
\item The original non-node sample versus the non-node sample shuffled by mass; 
\item The non-node sample shuffled by mass and $\dfil$ versus the non-node sample shuffled by mass. 
\end{enumerate} 
For the full sample, (i) evaluates the total SGB and (ii) evaluates the SGB that can be attributed to $\dnode$. For the non-node sample, (iii) evaluates the total SGB, and (iv) assesses the SGB that can be attributed to $\dfil$.

The results are shown in terms of the ratio between the statistics taken from the samples we are comparing. As we have discussed in \autoref{sec:strength}, any difference from unity in the ratio can be seen as a sign of SGB in the sample, and the errors are determined from jackknife subsamples of the simulation box, providing an estimate of the statistical importance of the signal. Although the results here are based on one random shuffle of each type, we have tested that our results are consistent regardless of the random seed used in the shuffling process.
We explore one aspect of the physics underlying our findings in \autoref{sec:appendixD}.

\subsubsection{Full sample and $\dnode$ effect}
\label{sec:sgb_signal_all_full}

\begin{figure*}
    \centering
    \includegraphics[width=0.85\textwidth, trim=1.6cm 0 2cm 0, clip]{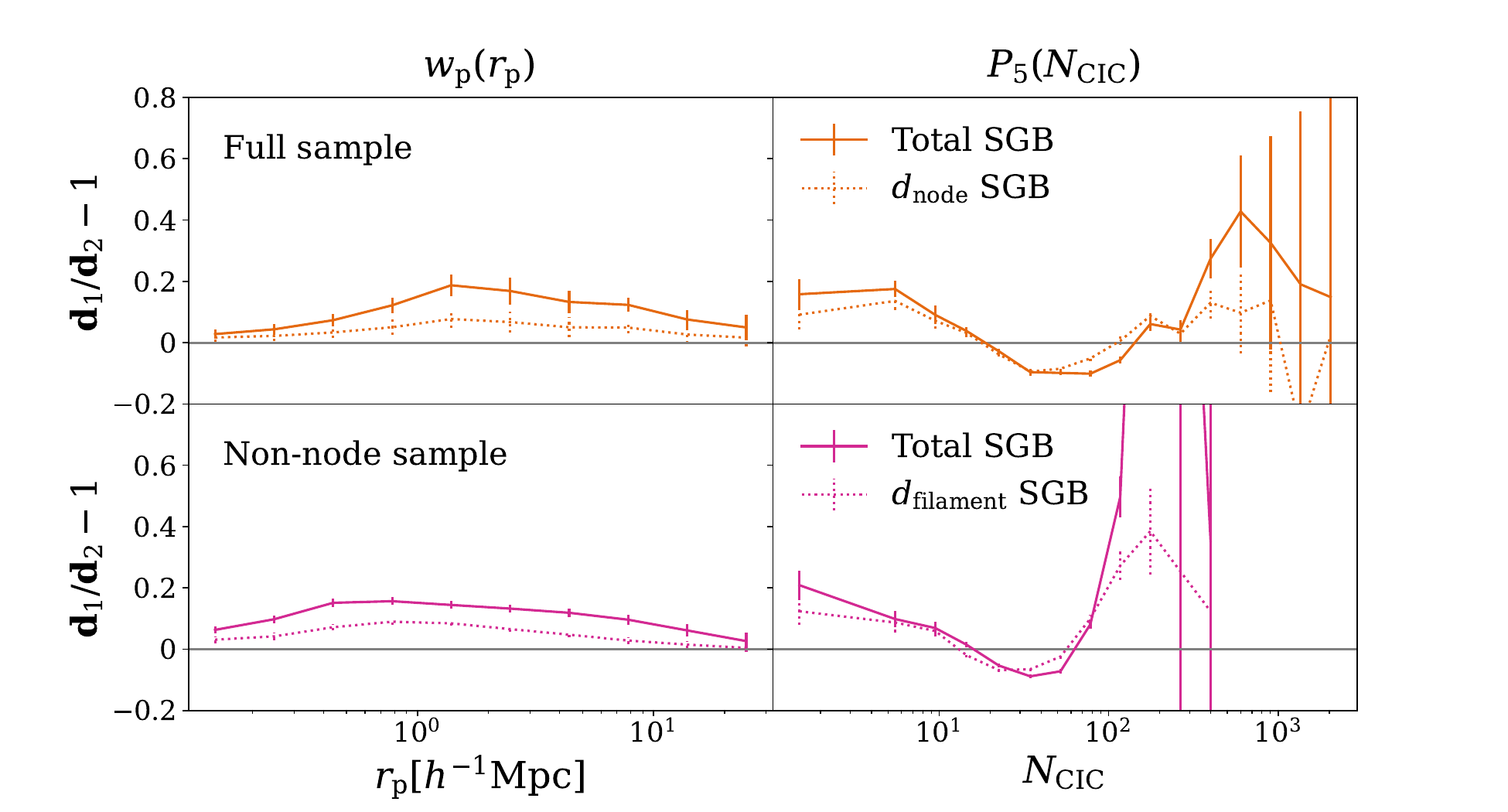}
    \caption{In this figure we show the relative contribution of the secondary galaxy bias (SGB) to the data vector, $\mathbf{d}$, of each statistic: $\wprp$ (left column) and $\Pncic$ (right column). 
    We show the environment specific SGB contribution in dotted lines, in comparison with the total SGB shown in solid lines.
    We plot the measured signals for both the full sample (top row) and the non-node sample (bottom row).
    The SGB signal is presented in terms of $\mathbf{d}_{\rm 1} / \mathbf{d}_{\rm 2}-1$, as defined in \autoref{sec:ratio}.
    Deviations of $\mathbf{d}_{\rm 1} / \mathbf{d}_{\rm 2}-1$ from the horizontal zero-line indicate an SGB signal where SGB either increases (positive $\mathbf{d}_{\rm 1} / \mathbf{d}_{\rm 2}-1$) or decreases (negative $\mathbf{d}_{\rm 1} / \mathbf{d}_{\rm 2}-1$) the statistic.
     We extract the total SGB signature from the ratio between the original and mass shuffled measurements, and the environment specific SGB signatures from the ratio between the double shuffled and mass shuffled measurements.
    This figure illustrates a statistically significant measurement of the node and filament contribution to the total SGB.}
    \label{fig:stat_comp}
\end{figure*}

We first examine the SGB in the full sample, along with the contribution of $\dnode$.
In the top row of \autoref{fig:stat_comp}, we show the results of comparisons (i) and (ii), which indicate the strength of the SGB present in the full sample. The solid curves with error bars represent comparison (i), the total SGB. We observe considerable SGB in the sample, as indicated by both $\wprp$ and $\Pncic$. The $\wprp$ results demonstrate that SGB increases clustering in the range of scales that we investigate. The most prominent effect is seen at intermediate scales, since the shuffling process preserves the 1-halo term in the clustering, diminishing the difference at small scales, while the underlying secondary halo bias weakens at large scales.

The $\Pncic$ results show that SGB increases the likelihood of having a large and a small number of companion galaxies, while reducing the proportion of intermediate companion counts. This suggests that the overall effect of SGB is that the more clustered haloes tend to host more galaxies, adding to the large groups of galaxies in the $\Ncic$ distribution. At the same time, the less clustered haloes host fewer galaxies, resulting in more empty space in the galaxy distribution, which is reflected in the increased probability of low $\Ncic$.

The dotted lines with error bars represent the SGB associated with $\dnode$ from comparison (ii). The comparison between the dotted and solid lines shows that $\dnode$ has an effect on galaxy occupation in the same way as the total effect, namely, haloes with lower values of $\dnode$ (which are closer to their neighbouring nodes) tend to host more galaxies at the same mass. This is in agreement with the results from \autoref{sec:hod_measure}. Although $\dnode$ contributes significantly to the total SGB, it is not the only factor. It is not possible to determine the exact amount of contribution from $\dnode$ due to scale dependences and the fact that the ratios do not translate directly to a physical fraction.

\subsubsection{Non-node sample and $\dfil$ effect}
\label{sec:sgb_signal_all_noncl}

We investigate the SGB in the non-node sample, from which node galaxies are excluded, and the contribution from $\dfil$. The bottom row of \autoref{fig:stat_comp} displays the results. The solid curves with error bars represent comparison (iii), the total SGB, and the dotted curves correspond to comparison (iv), the $\dfil$-related SGB. The signals we detect are similar to those in \autoref{sec:sgb_signal_all_full}, with $\wprp$ enhanced on all scales and $\Pncic$ increased for small and large companion counts. This implies that even when node galaxies are excluded, haloes that are more clustered tend to host more galaxies. For $\dfil$ in particular, more galaxies are found in haloes closer to the filaments, which is consistent with \autoref{sec:hod_measure}. The effect of $\dfil$ can explain a significant part of the total secondary bias, but not all of it.

The discrepancies between the full sample and the non-node sample are evident. The uncertainties in the ratios are lower for the latter, suggesting that the effect of SGB is less reliant on the environment when nodes are excluded, implying that in the extreme environment of nodes, galaxy occupation has more varied behaviour. The peak of the difference in $\wprp$ is on a slightly smaller scale than for the full sample, which is likely due to the smaller radii of haloes farther away from the nodes, and thus the earlier emergence of the 2-halo term. The effect of SGB on the small scale $\wprp$ is reduced by the exclusion of node regions, which is likely the cause of the small discontinuity at a few $\Mpch$. The $\Pncic$ measurements are cut off at a smaller $\Ncic$ (around $\Ncic\sim400$) for the non-node sample, due to the reduced group sizes without the node galaxies.
Note, we limit the figure range to emphasise the statistically significant behaviour in the non-node sample for $\Ncic$ below 100.
For the domain of larger $\Ncic$ values, the error bars are too large to identify any systematic behaviour that may be relevant.
While there may be other important scale-dependent effects in this domain, this data set does not have the appropriate sample to probe such an effect.

\section{Dependence of Secondary Galaxy Bias on Environment}
\label{sec:sgb_split}

In the preceding section, we have studied the relative contribution of environmental measures as secondary halo properties to the total SGB in a galaxy sample. In this section, we explore the role of the cosmic web environment in the SGB from a different angle: whether galaxy samples with similar halo mass distributions but different environments display different levels of SGB. It is well known that both the secondary halo bias and the secondary galaxy bias are sensitive to halo mass \citep[e.g.,][]{wechsler06,wang2022_SDSS_CIC}. This, combined with the fact that halo masses are strongly correlated with the environment, presents a challenge for our analysis. Therefore, when comparing the SGB in different cosmic web environments, we need to separate the effect of $\dnode$ or $\dfil$ from the effect of the halo mass. To accomplish this, we divide the galaxy sample at the 50th percentile of the host $\dnode$ or $\dfil$ within each narrow bin of host halo masses, instead of percentiles in the entire sample. This approach ensures that the split subsamples have similar distributions of halo masses and prevents the halo mass dependence of SGB from masquerading as a dependence on $\dnode$ or $\dfil$.

\begin{figure*}
    \centering
    \includegraphics[width=0.85\textwidth, trim=1.5cm 0.5cm 2cm 0cm]{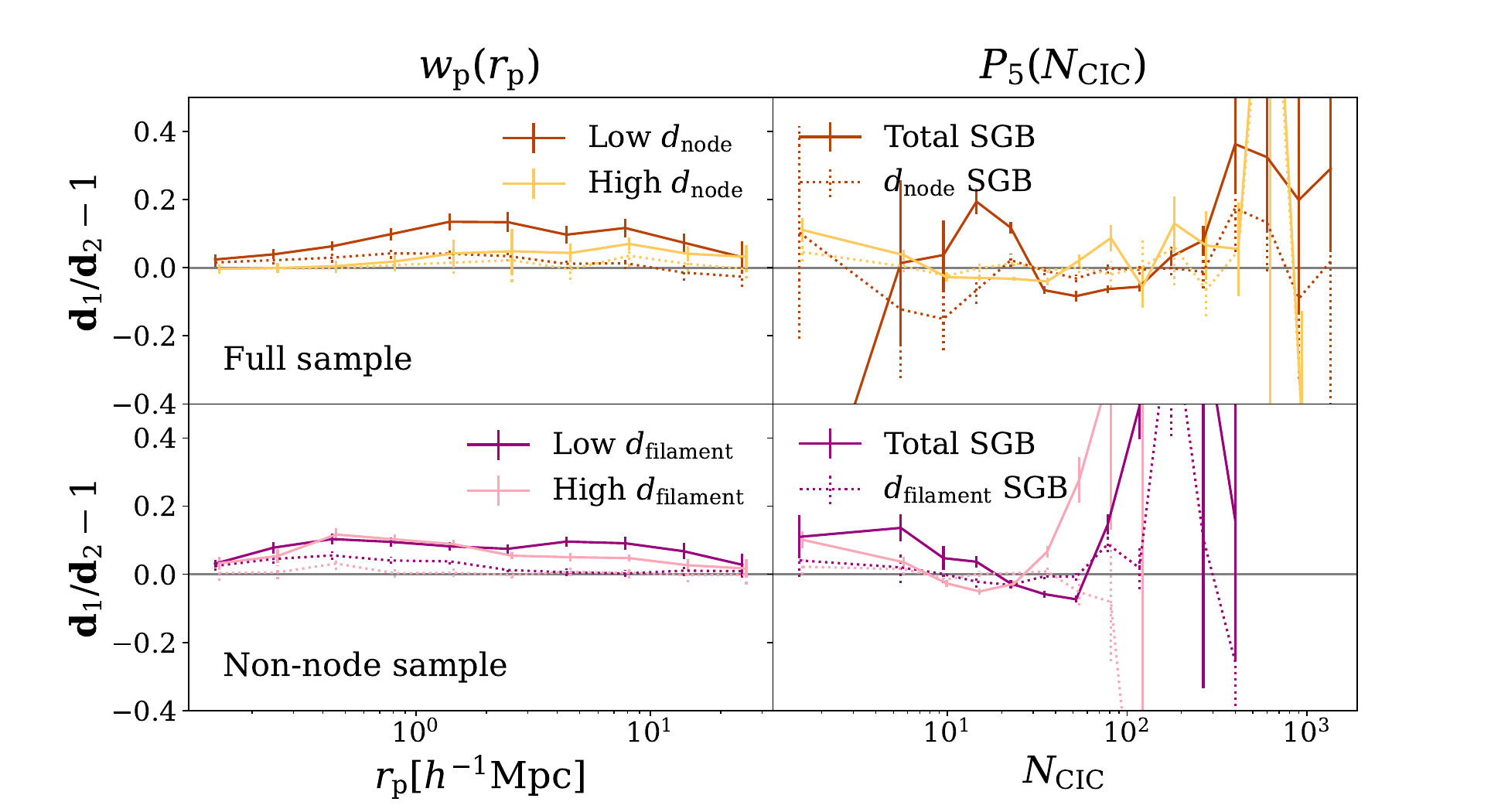}
    \caption{In this figure, we show the strength of the secondary galaxy bias (SGB) signal for galaxies in different node or filament environments.
    The solid and dotted curves correspond to the respective total and distance related SGB as in \autoref{fig:stat_comp}.  But, the different colours here correspond to subsamples for the low and high $\dnode$ or $\dfil$ subsamples separately, as labelled in the left column.  The total SGB is stronger for the subsample closer to nodes or filaments.  Within each subsample, the environmental component weakens.}
    \label{fig:mark_split}
\end{figure*}

\subsection{Secondary galaxy bias at different $\dnode$}
\label{sec:sgb_split_dcl}

We investigate the dependence of SGB on the node environment by comparing the SGB signals in the two galaxy subsamples with low and high $\dnode$. The upper row of \autoref{fig:mark_split} shows our measurements of the SGB effect of both statistics. The two subsamples with low and high $\dnode$ are represented by different colours, as labelled in the top left panel. The solid curves and the dotted curves correspond to the total SGB and the $\dnode$-related SGB, respectively, as labelled in the top right panel, similar to \autoref{fig:stat_comp}. We can see that the SGB signal is stronger in the high $\dnode$ subsample than in the low $\dnode$ subsample.

We find that for both subsamples, the total SGB increases the two-point clustering in the range of scales we investigate, and shifts companion counts in cylinders towards more extreme values\footnote{The only exception is in the lowest $\Ncic$ bin for the subsample closer to nodes, where the effect of the SGB reduces the probability, suggesting that the SGB in dense environments disfavours extreme isolation of galaxies.}, similar to the results from the entire sample. This can be interpreted as a positive correlation between halo clustering and galaxy occupation. In other words, in both subsamples, the haloes that are more strongly clustered also contain more galaxies.

By comparing the two subsamples, it is evident that the total SGB is significantly stronger at lower $\dnode$. The contrast in $\wprp$ between the two subsamples is most noticeable at small scales, where the low $\dnode$ subsample shows a positive signal, while the high $\dnode$ curve is consistent with zero. This can be explained by the shuffling procedure, which preserves the 1-halo term and the larger separations between haloes in environments farther away from nodes, resulting in the 2-halo term appearing at larger scales. For $\Pncic$, there is an overall decrease in counts in the high $\dnode$ subsample compared to the low $\dnode$ subsample, due to the lower number density of galaxies away from nodes, as well as the weakening of the SGB signal.

We now investigate the $\dnode$-related SGB, which is represented by the dotted lines. In the low $\dnode$ subsample, there is a weak $\dnode$-related SGB signal, while the high $\dnode$ subsample shows little evidence of $\dnode$-related SGB. This is in agreement with our findings in \autoref{sec:hod_measure}, which suggest that the HODs of samples further away from the nodes are less distinct from each other. Additionally, for both subsamples, the proportion of $\dnode$-related SGB to the total SGB in the subsamples is lower than in the entire sample, indicating that the dependence of galaxy occupation on $\dnode$ is largely explained by the coarse division of galaxies into low and high $\dnode$ subsamples. This also shows that the signal of $\dnode$ and $\dfil$-related in \autoref{fig:stat_comp} can be largely ascribed to galaxy pairs across different environments.

\subsection{Secondary galaxy bias at different $\dfil$}
\label{sec:sgb_split_dfil}

We investigate the dependence of the SGB on $\dfil$ in the non-node sample. The bottom row of \autoref{fig:mark_split} displays the results. The $\dfil$-related SGB effect is similar to the $\dnode$ effect discussed in \autoref{sec:sgb_split_dcl}, with the total SGB being stronger at lower $\dfil$ than higher $\dfil$, although the difference is not as pronounced as in the $\dnode$ case. The $\wprp$ statistic reveals a weak signal of $\dfil$-related SGB in the low $\dfil$ subsample, while $\Pncic$ hardly shows a signal. In contrast, neither statistic detects a strong $\dfil$-related SGB in the high $\dfil$ subsample.

In summary, regardless of whether the sample is divided into low or high $\dfil$, it is evident that more clustered haloes contain more galaxies. However, the preference is more pronounced in the low $\dfil$ subsample. Furthermore, the amount of $\dfil$-related SGB is significantly reduced in both subsamples after the splitting, implying that the number of galaxies is mainly determined by the general type of environment in relation to nearby filaments, rather than by minor $\dfil$ variations.

\section{Discussion}
\label{sec:discussion}

In this study, we investigated the impact of the cosmic web on the SGB by examining its effect on galaxy clustering measurements. We will now discuss the implications of our findings, as well as some of the restrictions of this study.

In \autoref{sec:hod_measure}, we compared the halo occupation distribution in different environments. We found that at fixed halo mass, haloes close to nodes and filaments host more galaxies. Rather than parametrised fits, this halo occupation distribution measurement is a direct measurement of the SGB attributable to the distance between the host halo and dense cosmic web structures. Early works \citep{einasto2003,einasto2005} found that haloes in denser environments, as measured by the distance to close neighbours, are richer in their galaxy content. More recent studies, for example, \citet{croft2012,zehavi2018} and \citet{Bose_2019}, have also found that the halo occupation distribution is higher for haloes in environments with higher intermediate-scale overdensities. These findings are broadly consistent with ours, although we use different proxies for the environment compared to the overdensity criteria used in these works: haloes located near nodes and filaments tend to have more close neighbours, as well as surrounding overdensities higher than those of other haloes, .

Our research is one of the first to investigate the role of the cosmic web in the secondary galaxy bias effect. \citet{hadzhiyska2020} used the \textsc{IllustrisTNG} simulation to explore the effect of local environment by employing a proxy of the local mass density and found that galaxies in similar environments tend to cluster together, which is in line with our results that the coarse division of the galaxy sample by types largely explains the environment-related secondary galaxy bias. \citet{Xu2021} studied the relative contribution of cosmic web environment types to the total secondary galaxy bias using the \textsc{Millennium} simulation and concluded that the environment type measured on scales of $5-10\Mpch$ constitutes a considerable portion of the total secondary bias signature. This is in agreement with our findings from \autoref{sec:sgb_denv}, although we use different indicators of the environment.
As we were completing this manuscript, we became aware of an independent analysis by \citet{montero-dorta2023}, who also used distances from cosmic web structures to describe the environment.
They found that at fixed halo mass, objects closer to dense structures cluster more strongly, which accounts for a significant portion of the dependence of galaxy clustering on halo formation time, also in qualitatively agreement with our findings.
More recently, \citet{perez2024} compared between HODs in different cosmic web environments in \textsc{IllustrisTNG}.
While finding less dependence of the HOD on the filamentary environment, they found a significant excess of faint galaxies at lower halo masses near nodes, in broad consistency with our results.

In \autoref{sec:sgb_split}, we present a novel element in the relation between the cosmic web and the SGB: to what extent haloes in different cosmic web environments exhibit different SGB behaviours. We find that haloes close to nodes and filaments are subject to stronger SGB. We argue that this is an important component of the cosmic web effect, as it sheds light on fundamental differences in the physics of galaxy formation and evolution between different environments.

In this work, we have focused on the connection between the cosmic web and the SGB, in other words, the response of the galaxy–halo connection to the environment. We note that there is a relatively larger volume of work on the influence of the environment on the halo bias. For example, \citet{pujol2017} and \citet{shi_sheth2018} claimed that halo clustering is completely determined by the local environment. \citet{paranjape18} found that haloes in isotropic and anisotropic environments show different halo assembly biases, \citet{ramakrishnan2019} showed that halo clustering depends on internal halo properties only through tidal anisotropy, and \citet{mansfield2020} proposed the tidal and gravitational effects of the surrounding large-scale structure as main causes of low-mass halo assembly bias. These results indicate that the environments of haloes play a physically fundamental role in determining the halo clustering, which is connected to the traditionally studied halo assembly bias.

We discuss how the cosmic web effect on SGB connects to some of the more commonly studied secondary halo properties. In particular, studies have extensively examined the secondary galaxy bias associated with halo concentration, formation time, spin, etc. Each of these halo properties affects the galaxy occupation beyond halo mass \citep[see, e.g.,][for a systematic study]{Xu2021}.  Haloes in different cosmic web environments have systematically different assembly histories that are reflected in their secondary properties.  For example, haloes that frequently merge are likely to have later formation times and lower concentrations. It has also been shown that low-concentration haloes tend to host more satellite galaxies \citep[e.g.,][]{wang2022_SDSS_CIC}, consistent with late formers having more frequent recent mergers.  Although the cosmic web and traditional secondary properties are connected, we argue that the cosmic web environment provides a more fundamental view of the factors that affect galaxy formation and evolution. 

The cosmic web descriptors are linked to the causal elements of the assembly histories. For instance, node haloes often experience frequent mergers, which are supplied by the filaments that connect them. Moreover, haloes in different cosmic web environments experience different tidal fields. At the most extreme end of the environmental range, massive node haloes have a major influence on their tidal environment and affect nearby haloes through anisotropic tidal forces. These cosmic web descriptors include distances to dense structures, which are used in this work, and measurements of the surrounding density, which are used in other works. By using cosmic web descriptors as a secondary feature, we can investigate their role in the formation of galaxies.

In this work, we study the secondary galaxy bias, which explicitly excludes the effect of halo mass on galaxy properties, and we underline the importance of disentangling the halo mass effect from the contribution of any secondary factor to galaxy formation and evolution. The success of various galaxy–halo connection models \citep[see][and references therein]{wechsler_tinker18} has demonstrated that halo mass (or some mass-like measure) is the predominant determinant of the properties of its galaxies. As halo mass is known to correlate with almost all other halo properties \citep[e.g.,][]{wechsler02,maccio2007,knebe2008}, any apparent sensitivity of galaxy properties to secondary halo properties could, in fact, have a root in the halo mass dependence. It is crucial to always account for halo mass in the theoretical framework, and while it is more challenging to estimate halo masses in observational data, careful considerations of its effect should be made before drawing conclusions on physical factors that impact galaxy formation and evolution.

One might posit that any environmental dependence of galaxy occupation might be due to a halo mass dependence: we expect more massive haloes to prefer overdense regions of our Universe. However, our research has revealed that galaxies prefer to live near nodes and filaments, even when the halo mass is taken into consideration.  This preference indicates that the cosmic web has a more complex effect on galaxy physics. These effects could be due to the different halo assembly histories, as well as surrounding tidal anisotropies, which we have discussed above. Galaxies in haloes with different assembly histories will form in different potential wells and have different merger histories, leading to different star formation histories, dynamical states and morphologies. On the other hand, the anisotropic tidal field may strip galaxies of their cold gas, or heat the gas reservoir, thus suppressing star formation as galaxies move through the cosmic web \citep[e.g.,][]{guo2021,guo2023}.

There have been numerous observational works on the dependence of the halo occupation on factors beyond halo mass, investigating different secondary properties, and obtaining mixed results \citep[e.g.,][etc.]{lin_mandelbaum_etal15,miyatake_etal16,vakili_2019,salcedo2022}.
Relatively few papers have observationally studied the role of the environment in this context, however also with mixed results.
\citet{paranjape18} found in SDSS DR7 that the effect of the tidal environment on halo occupation is likely to be weak.
On the other hand, \citet{alfaro2022} found in SDSS DR12 that the halo occupation is significantly increased in node-like environments, and significantly decreased in voids, in better consistency with our findings.
More recently, \citet{paviot2024} found that luminous red galaxies in eBOSS prefer denser and more anisotropic environments.
\citet{alam2024} found significant dependence of galaxy occupation on the tidal environment of host haloes in GAMA.
\citet{yuan2024} found hints of environment dependence of galaxy occupation in haloes in the DESI One Percent Survey.

Our analyses in the main text are based on galaxies with stellar masses above $10^8M_\odot$ in TNG300-1, which has a dark matter particle resolution of $4\times10^7\Msunh$.
As a result, some of the less massive haloes that host these galaxies contain of order $10^3$ dark matter particles.
To test the robustness of our qualitative results to this resolution limit, we repeat the analysis in \autoref{sec:sgb_signal_all} for galaxy samples with stellar mass thresholds of $10^9M_\odot$ and $10^{10}M_\odot$, which correspond to more massive haloes.
We show the results in \autoref{sec:appendixA}, and find they are qualitatively similar to the $10^8M_\odot$ sample.
This suggests that the resolution effect does not significantly affect our results, in consistency with the expectations from \citet{mansfield_avestruz2021}, who showed that most halo properties are converged at $10^3$ particles, for host haloes and subhaloes alike.
Therefore, we argue that our qualitative findings are not sensitive to the resolution limit of the simulation.

Our findings are based on the cosmic web structure identified by the \texttt{DisPerSE} cosmic web finder. Other algorithms, such as those discussed in summarised in Table~1 of \cite{libeskind2018}, may lead to different descriptions of the environment of individual objects. Nevertheless, the general behaviour of these algorithms is in agreement with each other, and we do not anticipate our primary conclusions to be altered by alternative cosmic web identification methods. It is worth noting that our quantification of the environment of haloes and galaxies, i.e., the distance to nearby dense structures, is not a comprehensive description of the environment information. For instance, this metric does not take into account the relative location of an object along a filament, nor does it differentiate between nodes or filaments with different densities and sizes. We do not consider cosmic sheets and voids in this work either. Therefore, we cannot definitively rule out the possibility that the secondary galaxy bias is completely rooted in the cosmic web environment.

Our analysis demonstrates the ability of $\Pncic$ to investigate the nuances of secondary galaxy bias, with a statistically significant measurement of environmental contributions to the SGB. As argued in \citet{wang2022_SDSS_CIC}, while the two-point correlation function mainly concentrates on the densest parts of the galaxy distribution, the counts-in-cylinders statistic, $\Pncic$, is sensitive to all but the most extreme underdensities, and measures higher-order statistics of the galaxy field. In forward modelling approaches, $\Pncic$ provides additional information on the two-point statistics, and in our shuffling procedure, the changes in $\Pncic$ also reveal a level of detail that contributes to our understanding of the underlying physics.
In both \autoref{sec:sgb_denv} and \autoref{sec:sgb_split}, we observe that the SGB affects $\wprp$ and $\Pncic$ in different manners.
Specifically, the SGB effect causes an overall increase in the amplitude of $\wprp$, whereas it drives the distribution of $\Ncic$ towards more extreme values, producing a non-trivial change in the shape of $\Pncic$.
Measurements of this shape dependence has the potential to break degeneracies between the SGB and other factors that affect galaxy clustering, for example, halo clustering.
These findings illustrate that different probes reveal different aspects of the underlying physics, and provide an avenue to exploit existing data more efficiently.

\section{Conclusions}
\label{sec:conclusion}

In this study, we explore a link between the cosmic web environment and the galaxy–halo connection. 
First, we treat the host halo proximity to nodes and filaments as a secondary halo property, and quantify its relative contribution to the total secondary galaxy bias.
Second, we compare the behaviour of the secondary galaxy bias in different environments.
Our findings are summarised as follows.
\begin{itemize}

    \item We identify dense structures in the cosmic web (nodes and filaments) in the TNG300-1 run of the \textsc{IllustrisTNG} simulation, using the \texttt{DisPerSE} algorithm.
    We use halo distances to these dense structures as an environmental measure.
    We illustrate general features of the cosmic web with these measures in \autoref{fig:gal_scatter_dis} and \autoref{fig:gal_fraction_dis}.

    \item We directly measure the halo occupation distribution for our galaxy sample with stellar masses above $10^8\Msun$, and find that haloes closer to nodes or filaments tend to host more galaxies at fixed halo mass (\autoref{fig:hod}).
    
    \item We compare summary statistics of shuffled and original galaxy samples to quantify the total secondary galaxy bias and the component that can be attributed to our environmental measures (see \autoref{fig:shuffling_cartoon} for a schematic illustration). In addition to the projected two-point correlation function, $\wprp$, we include a novel perspective with the counts-in-cylinders statistics, $\Pncic$ (see \autoref{fig:CIC_cartoon} for the definition of $\Pncic$).
    \autoref{fig:stat_orig} provides examples of both statistics.
    
    \item In our chosen summary statistics, we confirm that the secondary galaxy bias causes an enhancement in the two-point clustering, and we expose a nuanced effect with the counts-in-cylinders statistics, which manifests as a redistribution of galaxies from intermediate sized companion groups into groups with either very large or very small numbers of galaxies (solid curves in \autoref{fig:stat_comp}).
    We conclude that the total effect of secondary galaxy bias is for galaxies to preferentially reside in more strongly clustered haloes at similar halo masses. 

    \item We find that the host halo distance to nodes or filaments can account for a significant portion of the total secondary galaxy bias, but not the entire effect (see comparison between solid and dotted curves in \autoref{fig:stat_comp}).

    \item We find that the total effect of secondary galaxy bias is relatively stronger for subsamples that are closer to nodes or filaments (see comparison between different coloured solid curves in \autoref{fig:mark_split}). Within each subsample, the environmental component of SGB weakens (see reduced deviation of the ratio from unity in dotted curves in \autoref{fig:mark_split}). This trend indicates that while host haloes closer or further from dense cosmic web structures have different galaxy occupations, the finer details of the environment beyond this qualitative classification is less important in the galaxy–halo connection. We stress the importance of comparing galaxy samples with similar host halo mass distributions to isolate the effects of environment.

\end{itemize}

This work lays out a framework to comprehensively investigate the role of the cosmic web in the galaxy–halo connection, and constitutes a critical step towards understanding the role of the environment in galaxy formation and evolution.
In the future, we will explore alternative descriptions of the cosmic web, and extend the analysis to observational data.

\section*{Acknowledgements}

We thank Johannes Lange, Risa Wechsler, Andrew Zentner, and Qiong Zhang for useful discussions.
We thank the referee for their comments and suggestions that helped improve the manuscript.

KW acknowledges support from the Leinweber Postdoctoral Research Fellowship at the University of Michigan.  CA acknowledges support from the Leinweber Center for Theoretical Physics and DOE grant DE-
SC009193. HG is supported by the National SKA Program of China (grant No. 2020SKA0110100), National Natural Science Foundation of China (Nos. 11922305, 11833005), the CAS Project for Young Scientists in Basic Research (No. YSBR-092) and the science research grants from the China Manned Space Project with NOs. CMS-CSST-2021-A02. PW is sponsored by Shanghai Pujiang Program(No. 22PJ1415100).

This research made use of Python, along with many community-developed or maintained software packages, including
IPython \citep{ipython},
Jupyter (\https{jupyter.org}),
Matplotlib \citep{matplotlib},
NumPy \citep{numpy},
SciPy \citep{scipy},
and Astropy \citep{astropy}.
This research made use of NASA's Astrophysics Data System for bibliographic information.

\section*{Data Availability}

The simulation underlying this article were accessed from publicly available sources: \url{https://www.tng-project.org/data/}.
The catalogues including the cosmic web information will be shared on reasonable request to the corresponding authors.
The additional derived data are available in the article.


\bibliographystyle{mnras}
\bibliography{main,software}

\begin{appendices}

\counterwithin{figure}{section}
\counterwithin{table}{section}

\section{Alternative Stellar Mass Samples}
\label{sec:appendixA}

\begin{figure*}
     \centering
     \begin{subfigure}[b]{\textwidth}
         \centering
         \includegraphics[width=0.85\textwidth, trim=1.6cm 0 2cm 0, clip]{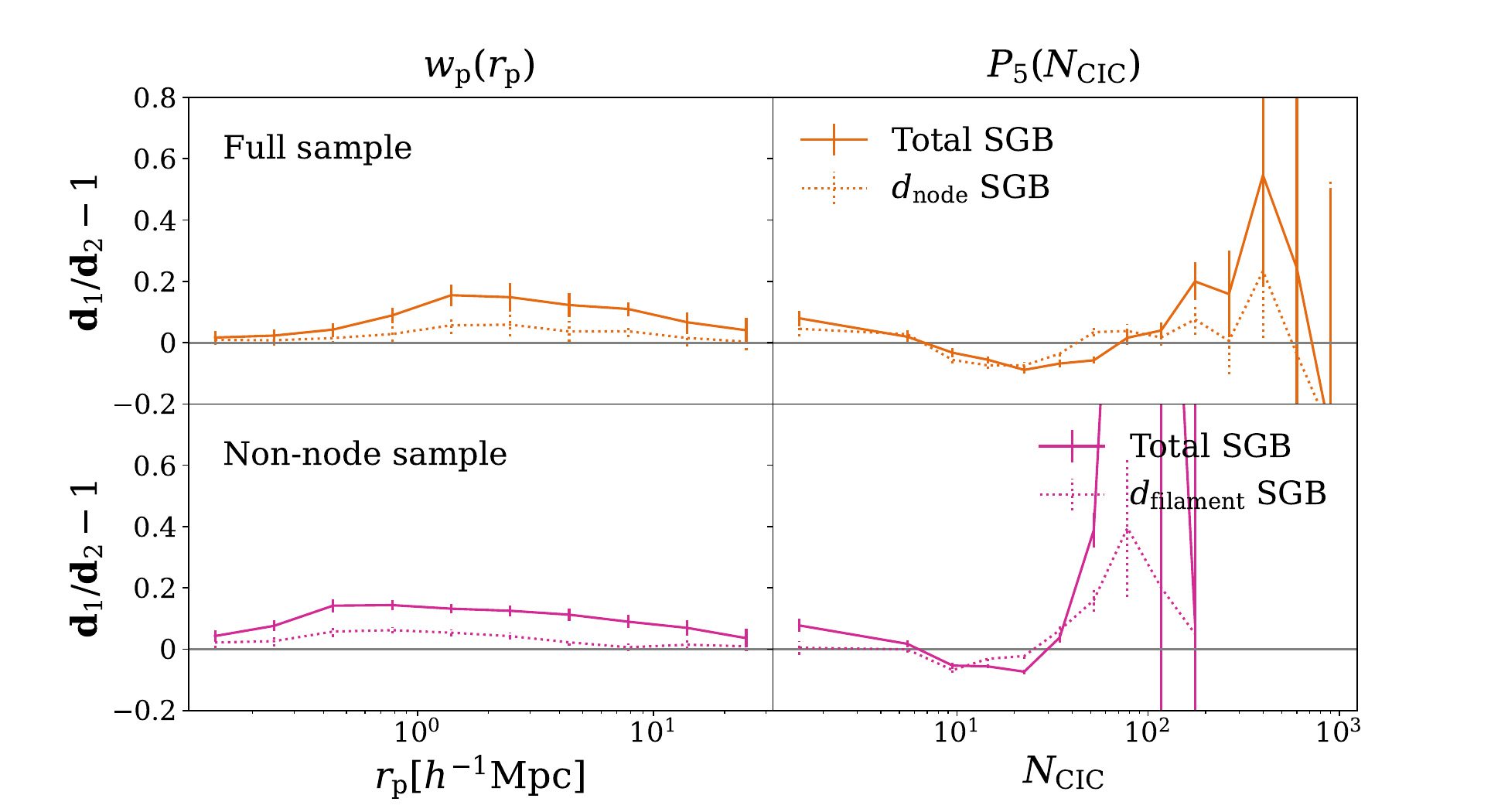}
         \caption{SGB signals for galaxies with stellar masses above $10^9\Msun$.}
         \label{fig:stat_comp_1e9}
     \end{subfigure}
     \hfill
     \begin{subfigure}[b]{\textwidth}
         \centering
         \includegraphics[width=0.85\textwidth, trim=1.6cm 0 2cm 0, clip]{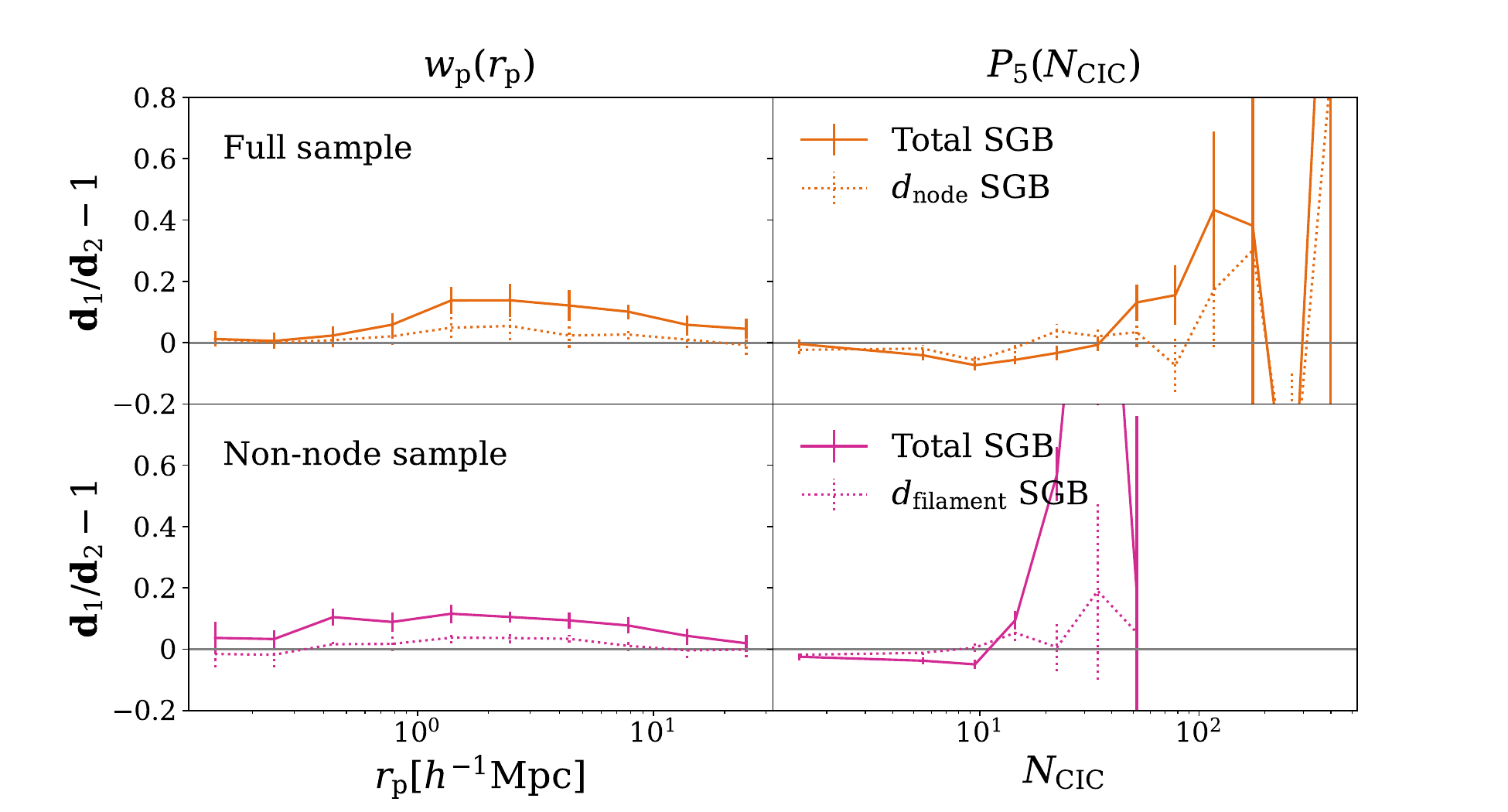}
         \caption{SGB signals for galaxies with stellar masses above $10^{10}\Msun$.}
         \label{fig:stat_comp_1e10}
     \end{subfigure}
     \caption{Same as \autoref{fig:stat_comp}, but for galaxy samples with stellar masses above $10^9\Msun$ (\autoref{fig:stat_comp_1e9}) and $10^{10}\Msun$ (\autoref{fig:stat_comp_1e10}) respectively. The results for these galaxy samples are qualitative similar to the $10^8\Msun$ sample in the main text.}
        \label{fig:stat_comp_1e9_1e10}
\end{figure*}

In this appendix, we repeat our analyses for alternative galaxy samples, with stellar mass thresholds of $10^9\Msun$ and $10^{10}\Msun$, instead of $10^8\Msun$ in the main text.
We limit our tests to these thresholds, because as can be seen in \autoref{fig:gal_fraction_dis}, samples with higher stellar masses are predominantly found very close to nodes and filaments, and have less different environments.
We show the results for these alternative samples in \autoref{fig:stat_comp_1e9} and \autoref{fig:stat_comp_1e10}.
Each figure is analogous to \autoref{fig:stat_comp}, where the two columns show results from $\wprp$ and $\Pncic$ respectively, and the two rows show results for the full and non-node samples respectively.
The solid curves indicate the strength of the total SGB from all sources, and the dotted curves are specific to $\dnode$ or $\dfil$-related SGB.
We find that for the more massive galaxy samples, the same conclusions hold as for the $10^8\Msun$ sample.
Namely, the effect of the total SGB is a preference of galaxies to populate haloes that are more strongly clustered; for $\dnode$ and $\dfil$, galaxies prefer haloes closer to the overdense structures, which effect accounts for a significant fraction of the total SGB, but does not explain all of it.
The only notable difference between the samples is that the signals in $\Pncic$ are shifted towards lower counts as the stellar mass threshold increases, because of the brighter samples have lower number densities and therefore fewer companions.
Because of this shift, the increase of $\Pncic$ at low counts due to SGB disappears in the $10^{10}\Msun$ sample.

\section{Alternative Cylinder Sizes}
\label{sec:appendixB}

\begin{figure*}
    \centering
    \includegraphics[width=0.95\textwidth, trim={1.8cm 0 3cm 0}, clip]{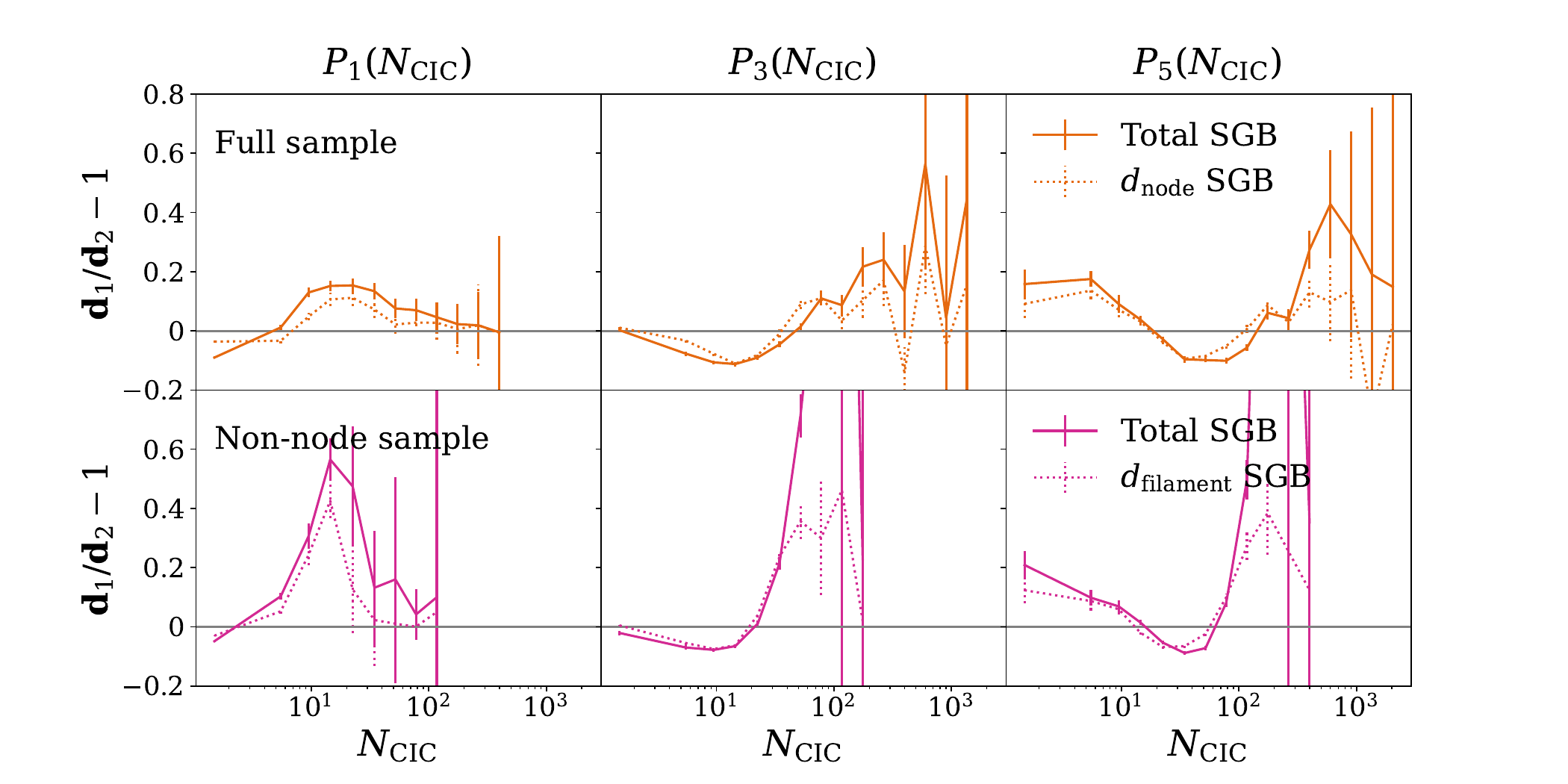}
    \caption{Each column illustrates the SGB signal measured with $\Pncic$, similar to the right column of \autoref{fig:stat_comp}, but for $\Pncic$ with different cylinder sizes (radius and half-length), 1, 3, and 5 $\Mpch$. The alternative cylinder size measurements are consistent with the results from the 5 $\Mpch$ measurements in the main text.}
    \label{fig:cyl_sizes}
\end{figure*}

Our main results are based on $\Pncic$ measured with the cylinder radius and half-length of 5 $\Mpch$.
In this appendix, we test the robustness of our conclusions by measuring $\Pncic$ with alternative cylinder sizes.
Our test results are shown in \autoref{fig:cyl_sizes}.
Each column in this figure is analogous to the right column of \autoref{fig:stat_comp}, but shows measurements with cylinder sizes of 1, 3, and 5 $\Mpch$ respectively, from left to right.
As can be naively expected, with smaller cylinder sizes, $\Pncic$ probes the more immediate surroundings of galaxies, and has generally fewer companions.
We observe an overall downscaling of $\Ncic$ with smaller cylinders compared to the 5 $\Mpch$ case.
For example, in the middle column, the qualitative behaviour of $\Pncic$ with the cylinder size of 3 $\Mpch$ is similar to the 5 $\Mpch$ measurement at $\Ncic\gtrsim20$, and the signal in the left column is similar to the 5 $\Mpch$ measurement at $\Ncic\gtrsim70$.
This is consistent with our main conclusions.
The effect of the SGB places more galaxies in highly clustered haloes, and leaves the less clustered haloes underpopulated.
With the large 5 $\Mpch$ cylinders, this increases the probability of having cylinders with both high $\Ncic$ in dense regions and low $\Ncic$ in underdense regions, as discussed in the main text.
However, the haloes in underdense environments that host very few galaxies are not probed by the smaller cylinders, which eliminates the increase of low $\Ncic$ probabilities.
The comparison between different cylinder sizes shows that we need to probe sufficiently large scales in order to fully observe the imprint of the SGB.

\section{Alternative Halo Mass Definition}
\label{sec:appendixC}

Our main results are obtained with the virial definition of halo mass, and in this appendix, we test the robustness of our qualitative findings with alternative halo mass definitions provided in the TNG group catalogues.
Instead of the virial mass, we use $M_{\rm 200m}$, the mass within a boundary that encloses a mean density equal to 200 times the mean density of the universe, and $M_{\rm 500c}$, which corresponds to 500 times the critical density of the universe.
Specifically, we use these alternative masses in the shuffling procedure, and shuffle among haloes that have similar $M_{\rm 200m}$ or $M_{\rm 500c}$.

\begin{figure*}
    \centering
    \includegraphics[width=0.85\textwidth, trim=1.6cm 0 2cm 0, clip]{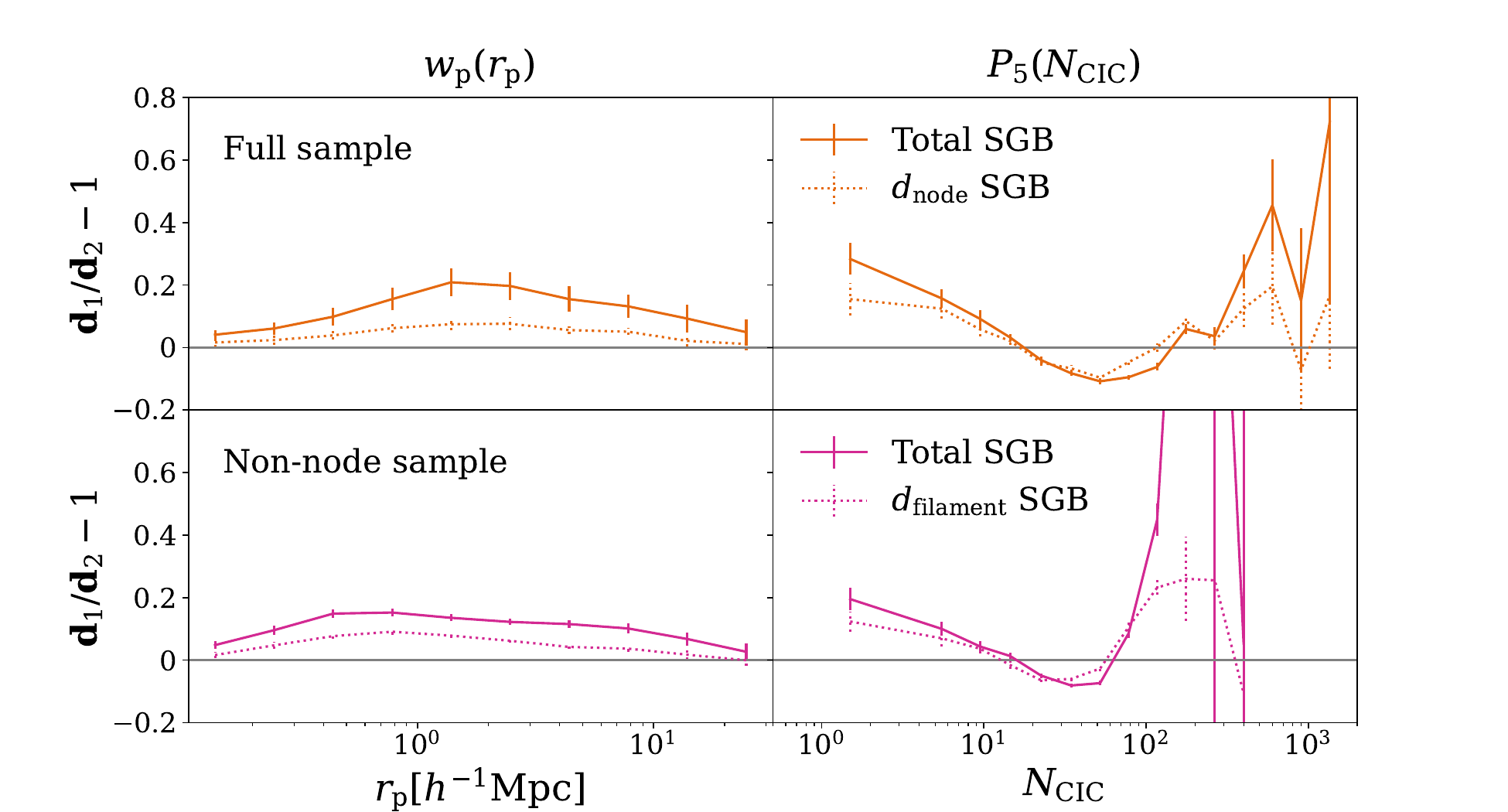}
    \caption{Similar to \autoref{fig:stat_comp}, but the shuffling is performed with $M_{\rm 200m}$, the halo mass within a boundary that encloses a mean density equal to 200 times the mean density of the universe.}
    \label{fig:stat_comp_M200m}
\end{figure*}

\begin{figure*}
    \centering
    \includegraphics[width=0.85\textwidth, trim=1.6cm 0 2cm 0, clip]{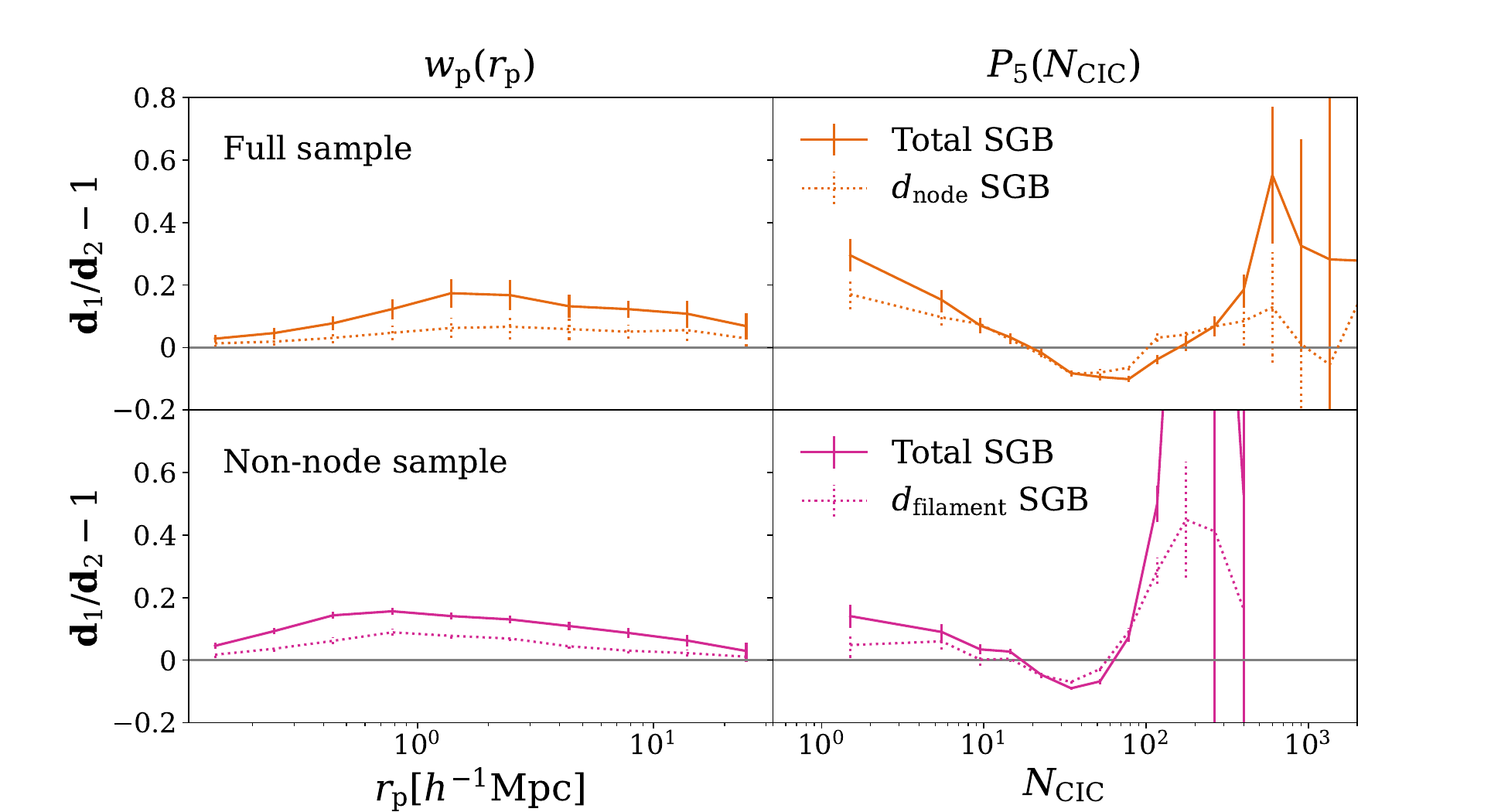}
    \caption{Similar to \autoref{fig:stat_comp}, but the shuffling is performed with $M_{\rm 500c}$, the halo mass within a boundary that encloses a mean density equal to 500 times the critical density of the universe.}
    \label{fig:stat_comp_M500c}
\end{figure*}

By comparing between \autoref{fig:stat_comp_M200m}, \autoref{fig:stat_comp_M500c}, and \autoref{fig:stat_comp}, we find that the results with alternative mass definitions are very similar.
This shows that our qualitative findings are not sensitive to the mass definition used for shuffling.

\section{Environment Dependence of Host Halo Gas Metallicity}
\label{sec:appendixD}

Higher gas metallicities in haloes indicate more star formation activity in the past.
In this appendix, we explore the dependence of the gas metallicity, $Z_{\rm gas}$, in host haloes on the distance from nearby nodes and filaments, to provide insight into our findings that galaxies prefer to populate haloes closer to these cosmic web structures.

We perform a mark correlation analysis between $Z_{\rm gas}$ and $d_{\rm node}$ (or $d_{\rm filament}$), which extracts the correlation between the properties while excluding their dependence on mass.
To do this, we split host haloes more massive than $10^{10}h^{-1}M_\odot$ into narrow mass bins of 0.1 dex, and within each mass bin, we calculate the percentile ranks --- marks --- of $Z_{\rm gas}$, $d_{\rm node}$ and $d_{\rm filament}$. 
We denote the marks by $\mathcal{M}(Z_{\rm gas})$, $\mathcal{M}(d_{\rm node})$, and $\mathcal{M}(d_{\rm filament})$.
We then calculate the mark correlation coefficient between the gas metallicity and the distances.
We find that for host haloes above $10^{10}h^{-1}M_\odot$,
\begin{equation}
   corr\left\{\mathcal{M}(Z_{\rm gas}),\mathcal{M}(d_{\rm node})\right\}=-0.32,
\end{equation}
and for the non-node host haloes above $10^{10}h^{-1}M_\odot$,
\begin{equation}
    corr\left\{\mathcal{M}(Z_{\rm gas}),\mathcal{M}(d_{\rm filament})\right\}=-0.24.
\end{equation}
The negative correlation coefficients show that, at the same halo mass, gas metallicity is higher for haloes closer to nodes or filaments.
This suggests that haloes closer to nodes or filaments have had more extensive star formation activity in the past, which likely contributes to the preference of these haloes by galaxies.

\end{appendices}

\label{lastpage}
\end{document}